\documentclass[12pt, a4paper]{article}
\usepackage{graphicx}
\usepackage{subeqnarray}
\begin{document}
%
%
\newenvironment{lefteqnarray}{\arraycolsep=0pt\begin{eqnarray}}
{\end{eqnarray}\protect\aftergroup\ignorespaces}
\newenvironment{lefteqnarray*}{\arraycolsep=0pt\begin{eqnarray*}}
{\end{eqnarray*}\protect\aftergroup\ignorespaces}
\newenvironment{leftsubeqnarray}{\arraycolsep=0pt\begin{subeqnarray}}
{\end{subeqnarray}\protect\aftergroup\ignorespaces}
\newcommand{\appleq}{\stackrel{<}{\sim}}
\newcommand{\appgeq}{\stackrel{>}{\sim}}
\newcommand{\arcsinh}{\mathop{\rm arcsinh}\nolimits}
\newcommand{\arctg}{\mathop{\rm arctg}\nolimits}
\newcommand{\diff}{{\rm\,d}}
\newcommand{\displayfrac}[2]{\frac{\displaystyle #1}{\displaystyle #2}}
\newcommand{\Erfc}{\mathop{\rm Erfc}\nolimits}
\newcommand{\Int}{\mathop{\rm Int}\nolimits}
\newcommand{\Nint}{\mathop{\rm Nint}\nolimits}
\newcommand{\pprime}{{\prime\prime}}

\title{An application of the tensor virial theorem  \\
       to hole + vortex + bulge systems}
\author{{R. Caimmi\footnote{
{\it Astronomy Department, Padua Univ., Vicolo Osservatorio 2,
I-35122 Padova, Italy} 
email: roberto.caimmi@unipd.it
}
\phantom{agga}}}

\maketitle
\begin{quotation}
\section*{}
\begin{Large}
\begin{center}

Abstract

\end{center}
\end{Large}
\begin{small}


The tensor virial theorem for subsystems
is formulated for three-component
systems and further effort is devoted
to a special case where the inner subsystems
and the central region of the outer one are
homogeneous, the last surrounded by an
isothermal homeoid.   The virial equations
are explicitly written under the
additional restrictions: (i) similar and
similarly placed inner subsystems, and
(ii) spherical outer subsystem.   An
application is made to hole +
vortex + bulge systems, in the limit
of flattened inner subsystems, which
implies three virial equations in three
unknowns.   Using the Faber-Jackson relation,
$R_e\propto\sigma_0^2$,
the standard $M_{\rm H}$-$\sigma_0$
form $(M_{\rm H}\propto\sigma_0^4)$
is deduced from qualitative
considerations.   The projected
bulge velocity dispersion to
projected vortex velocity 
ratio, $\eta=(\sigma_{\rm B})_{33}/\{
[(v_{\rm V})_{qq}]^2+[(\sigma_{\rm V})_{qq}]^2\}^{1/2}$,
as a function of the fractional radius,
$y_{\rm BV}=R_{\rm B}/R_{\rm V}$,
and the fractional masses,
$m_{\rm BH}=M_{\rm B}/M_{\rm H}$ and
$m_{\rm BV}=M_{\rm B}/M_{\rm V}$,
is studied in the range of
interest, $0\le m_{\rm VH}=
M_{\rm V}/M_{\rm H}\le5$ (Escala,
2006) and $229\le m_{\rm BH}\le795$
(Marconi and Hunt, 2003), consistent
with observations.
The related curves appear to be
similar to Maxwell velocity distributions,
which implies a fixed value of $\eta$
below the maximum corresponds to two
different configurations: a compact bulge
on the left of the maximum, and an
extended bulge on the right.    All curves
lie very close one to
the other on the left of the maximum, and
parallel one to the other on the right.
On the other hand, fixed $m_{\rm BH}$ 
or $m_{\rm BV}$, and $y_{\rm BV}$, are
found to imply more massive bulges
passing from bottom to top along a vertical
line on the $({\sf O}y_{\rm BV}\eta)$ plane,
and vice versa.    The model is applied
to NGC 4374 and NGC 4486, taking the fractional
mass, $m_{\rm BH}$, and the fractional
radius, $y_{\rm BV}$, as
unknowns, and the bulge mass is inferred
and compared with results from different
methods.    In presence of a massive
vortex $(m_{\rm VH}=5)$, the hole mass
has to be reduced by a factor 2-3 with
respect to the case of a massless vortex,
to get the fit.   Finally, the assumptions of
homogeneous inner bulge and isotropic
stress tensor are discussed.

\noindent
{\it keywords - Black hole physics; galaxies: bulges. PACS codes - 98.62.Js
}

\end{small}
\end{quotation}

%

\section{Introduction}\label{intro}
The existence of large masses confined in a restricted
central region of galaxies (hereafter quoted as ``holes''),
is widely
supported by current high-resolution observations
(for a review see e.g., Ferrarese and Ford, 2005;
Merritt, 2006).   There is an amount of increasing
evidence, that compact objects at the centre of
(sufficiently large) galaxies are
supermassive black holes (e.g., Maoz, 1998; Miller,
2006).   In addition, the presence of self-gravitating
nuclear disks (with typically several hundreds parsecs
in radius; hereafter quoted as ``outer vortexes'')
seems necessary to continuously replenish
inner Keplerian accretion disks (hereafter quoted as
``inner vortexes''), allowing substantial hole mass growth
(e.g., Escala, 2006).   Hole growth can successfully
be modelled as a rapid succession of vortexes with
comparable mass, related to a merger event (King et
al., 2008).    Most of the gas residing in the outer,
self-gravitating disk, is expected to be either turned
into stars, or expelled by the first generation of
high-luminosity stars, on a rapid (almost dynamical)
time-scale (King et al., 2008).

In the following, the disk made of the inner vortex
and the outer (self-gravitating) vortex, shall be
quoted as the ``vortex''.   Masses up to about
one fifth the inner bulge (i.e. bulge stars
inside the isophote with equatorial
axes coinciding with the edge of the vortex)
have been deduced from observations (e.g.,
Downes and Solomon, 1998).   The occurrence
of vortexes is expected to be a common
feature during the evolution of galaxies
hosting holes.   Accordingly, central
regions of large galaxies may
be modelled as the superposition of
three main components: a central hole,
a surrounding vortex, and an embedding
bulge.

In this view, the tensor virial theorem
for subsystems (Brosche et al., 1983;
Caimmi et al., 1984; Caimmi and Secco,
1992; Caimmi, 2007) could be a useful
tool of investigation.   In the case
under discussion there are nine virial
equations, three for each subsystem.
In absence of a satisfactory theory
of quantum gravity, the mass distribution
within the hole cannot be described,
which reduces the number of virial
equations from nine to six.   The
further restriction to an axisymmetric
gravitational potential, implying an
axisymmetric matter distribution for
each subsystem, further reduces the
number of virial equations from six
to four.   The final restriction to
flattened vortex and hole configurations,
reduces the number of virial equations
from four to three.

Keeping in mind that the virial theorem
deals with global instead of local
properties of matter distributions,
simple even if unrealistic density
profiles can be used for investigating
main features and general trends.
To this aim, the hole and the vortex
are modelled as concentric and coaxial,
flattened, homogeneous spheroids, the
inner bulge as a homogeneous sphere,
and the outer bulge as an isothermal
spherical corona, which allows an
analytical formulation (Caimmi and
Secco, 2002).

To close the system of three equations,
the number of unknowns must be reduced
to three.   The variables are: three
mass-averaged square velocities, three
masses, and three equatorial semiaxes.
Hole mass and equatorial semiaxis can
be related via the expression of the
gravitational (or Schwartzschild)
radius (e.g., Landau and Lifchitz,
1966, Chap.\,XI, \S97).   Typical values
of bulge and vortex equatorial semiaxes,
can be deduced from observations (e.g.,
Downes and Solomon, 1998).   The vortex
to hole mass ratio cannot exceed a value
of about 5 (e.g., Escala, 2006), deduced
from observations (Downes and Solomon,
1998).   An empirical correlation seems
to exist between bulge and hole mass
(e.g., Marconi and Hunt, 2003).
Mass-averaged bulge square velocities
may safely be approximated by square
luminosity-weighted second moments of
the line-of-sight velocity distribution
within the half-light radius, which is
weakly dependent on the details of the
orbital distribution (Cappellari et al.,
2006).   In fact, a similar quantity is
also used for studying the correlation
between hole mass and bulge velocity
dispersion (e.g., Gebhardt et al., 2000).
Taking the above informations into due
account, the number of unknowns in the
virial equations may be reduced from nine
to three.

In addition, the combination of two
virial equations yields a relation
between mass-averaged velocity ratios,
mass ratios, and equatorial axis
ratios, regardless of the values
of mass-averaged velocities, masses,
and equatorial semiaxes.   For assigned
bulge to vortex and vortex to hole mass
ratios, and vortex to hole equatorial
axis ratio, the mass-averaged bulge to
vortex velocity ratio is a function of
the bulge to vortex equatorial axis ratio,
and the related trend can be studied.
In particular, the bulge mass can be
determined by the knowledge of other
parameters, and the related value can
be compared with results from different
methods.

The current paper is organized
as follows.   The virial theorem
for subsystems, with regard to
three-component systems, is
formulated in Sect.\,\ref{bate}.
The model and its application to
hole + vortex + bulge systems are
the subject of Sect.\,\ref{mode},
where the results are also
presented and discussed, with
regard to the special cases of
NGC 4374 (M84) and NGC 4486 (M87).
The conclusion is drawn in
Sect.\,\ref{conc}.

\section{Basic theory}
\label{bate}

A general theory of two-component matter
distributions has been exhaustively
treated in earlier papers (e.g.,
Caimmi and Secco, 1992; Caimmi and
Marmo, 2003; Caimmi and Valentinuzzi,
2008) and an interested reader is
addressed therein and in parent
investigations (MacMillan, 1930,
Chap.\,III, \S76; Limber, 1959;
Neutsch, 1979; Brosche et al.,
1983; Caimmi et al., 1984).
The related tensor virial theorem
is expressed by $3\times2$ independent
equations, each containing $2+(2-1)$
terms corresponding to the kinetic
energy and the self potential energy
of the subsystem under consideration,
and to the tidal potential energy
due to the gravitational potential
induced by the other subsystem,
respectively.

In dealing with $N$ subsystems,
the tensor virial theorem is
expressed by $3\times N$ independent
equations, each containing $2+(N-1)$
terms corresponding to the kinetic
energy and the self potential energy
of the subsystem under consideration,
and to the tidal potential energies
due to the gravitational potential
induced by the other subsystems,
respectively.   Accordingly, the
results found for two-component
systems can be used, by inserting
in the virial equations
$N-2$ additional terms, related to the
tidal potential energies of the extra
components.

To this aim, the formulation of earlier
attempts (Caimmi and Secco, 1992, 2002)
shall be used.   If not otherwise stated,
for sake of convenience, matter distributions
shall be conceived as continuous media
instead of discrete particle sets (e.g.,
Limber, 1959; Caimmi, 2007).   In the
following, general definitions and related
explicit expressions will be presented.
Readers mainly interested to an application
of astrophysical interest (hole + vortex
+ bulge systems) might directly go
to the next Section \ref{mode}.

\subsection{The tensor virial theorem for
subsystems}
\label{tvst}

Let $i$, $j$, $k$, denote the subsystems
of a three-component matter distribution.
Owing to the additivity of the gravitational
potential, the potential-energy tensor and
the potential energy may be cast into the
form:
\begin{leftsubeqnarray}
\slabel{eq:pota}
&& (E_{\rm pot})_{pq}=\sum_u[(E_u)_{\rm sel}]_{pq}+\sum_u\sum_v
(1-\delta_{uv})[(E_{uv})_{\rm int}]_{pq}~~; \\
\slabel{eq:potb}
&& E_{\rm pot}=\sum_u(E_u)_{\rm sel}+\sum_u\sum_v(1-\delta_{uv})(E_{uv})_
{\rm int}~~; \\
\slabel{eq:potc}
&& p=1,2,3~~;\quad q=1,2,3~~;\quad u=i,j,k~~;\quad
uv=ij, ik, jk, ji, ki, kj~~;\qquad
\label{seq:pot}
\end{leftsubeqnarray}
where $\delta_{uv}$ is the Kronecker symbol,
$[(E_u)_{\rm sel}]_{pq}$ and $(E_u)_{\rm sel}$
the self potential-energy tensor and the self
potential energy, respectively, 
$[(E_u)_{\rm int}]_{pq}$ and $(E_u)_{\rm int}$
the interaction potential-energy tensor and the
interaction potential energy, respectively.
The related explicit expressions are (e.g.,
Caimmi and Secco, 1992):
\begin{leftsubeqnarray}
\slabel{eq:sela}
&& [(E_u)_{\rm sel}]_{pq}=\int_{S_u}\rho_u(x_1,x_2,x_3)x_p\frac
{\partial{\cal V}_u}{\partial x_q}\diff^3S_u \nonumber \\
&& \phantom{[(E_u)_{\rm sel}]_{pq}}=-\frac12\int_{S_u}\rho_u(x_1,x_2,x_3)
[{\cal V}_u(x_1,x_2,x_3)]_{pq}\diff^3S_u~~; \\
\slabel{eq:selb}
&& (E_u)_{\rm sel}=\int_{S_u}\rho_u(x_1,x_2,x_3)\sum_{s=1}^3x_s\frac
{\partial{\cal V}_u}{\partial x_s}\diff^3S_u \nonumber \\
&& \phantom{(E_u)_{\rm sel}}=-\frac12\int_{S_u}\rho_u(x_1,x_2,x_3)
{\cal V}_u(x_1,x_2,x_3)\diff^3S_u~~;
\label{seq:sel}
\end{leftsubeqnarray}
\begin{leftsubeqnarray}
\slabel{eq:inta}
&& [(E_{uv})_{\rm int}]_{pq}=-\frac12\int_{S_u}\rho_u(x_1,x_2,x_3)
[{\cal V}_v(x_1,x_2,x_3)]_{pq}\diff^3S_u~~; \\
\slabel{eq:intb}
&& (E_{uv})_{\rm int}=-\frac12\int_{S_u}\rho_u(x_1,x_2,x_3)
{\cal V}_v(x_1,x_2,x_3)\diff^3S_u~~;
\label{seq:int}
\end{leftsubeqnarray}
where $S$ is the volume, $\rho$ the density,
${\cal V}_{pq}$ and ${\cal V}$ the
gravitational tensor potential and potential,
respectively (Chandrasekhar, 1969, Chap.\,2,
\S10).

The tensor and the scalar virial theorem for
a single subsystem within the tidal field
induced by the other ones, read:
\begin{leftsubeqnarray}
\slabel{eq:vira}
&& 2[(E_u)_{\rm kin}]_{pq}+[(E_u)_{\rm sel}]_{pq}+[(E_{uv})_{\rm tid}]_{pq}+
[(E_{uw})_{\rm tid}]_{pq}=0~~; \\
\slabel{eq:virb}
&& 2(E_u)_{\rm kin}+(E_u)_{\rm sel}+(E_{uv})_{\rm tid}+(E_{uw})_{\rm tid}=0
~~; \\
\slabel{eq:virc}
&& u=i,j,k~~;\qquad v=j,k,i~~;\qquad w=k,i,j~~;
\label{seq:vir}
\end{leftsubeqnarray}
where $[(E_u)_{\rm kin}]_{pq}$ and $(E_u)_{\rm kin}$
are the kinetic-energy tensor and the kinetic
energy, respectively, 
$[(E_{uv})_{\rm tid}]_{pq}$ and $(E_{uv})_{\rm tid}$
the tidal potential-energy tensor and the
tidal potential energy, respectively.
The related explicit expressions are
(e.g., Caimmi and Valentinuzzi, 2008):
\begin{leftsubeqnarray}
\slabel{eq:kina}
&& [(E_u)_{\rm kin}]_{pq}=\frac12\int_{S_u}\rho_u(x_1,x_2,x_3)(v_u)_p(v_u)_q
\diff^3S_u~~; \\
\slabel{eq:kinb}
&& (E_u)_{\rm kin}=\frac12\int_{S_u}\rho_u(x_1,x_2,x_3)\sum_{s=1}^3[(v_u)_s]^2
\diff^3S_u~~;
\label{seq:kin}
\end{leftsubeqnarray}
\begin{leftsubeqnarray}
\slabel{eq:tida}
&& [(E_{uv})_{\rm tid}]_{pq}=\int_{S_u}\rho_u(x_1,x_2,x_3)x_p\frac
{\partial{\cal V}_v}{\partial x_q}\diff^3S_u~~; \\
\slabel{eq:tidb}
&& (E_{uv})_{\rm tid}=\int_{S_u}\rho_u(x_1,x_2,x_3)\sum_{s=1}^3x_s\frac
{\partial{\cal V}_v}{\partial x_s}\diff^3S_u~~;
\label{seq:tid}
\end{leftsubeqnarray}
where $(v_u)_p(v_u)_q=\overline{(v_u)_p(v_u)_q}$
and $[(v_u)_s]^2=\overline{[(v_u)_s]^2}$
are arithmetic means calculated within
the infinitesimal volume element,
$\diff^3S_u=\diff x_1\diff x_2\diff x_3$,
placed at ${\sf P}(x_1,x_2,x_3)$ (Binney
and Tremaine, 1987, Chap.\,4, \S4.3).
The tidal potential energy may be
conceived as the virial of the $u$
component in connection with the
tidal field induced by the $v$
component (Brosche et al., 1983).

The tensor and the scalar virial
theorem, expressed by Eqs.\,(\ref
{eq:vira}) and (\ref{eq:virb}),
are the extension of earlier results
related to one-component (e.g.,
Chandrasekhar, 1969, Chap.\,II, \S11;
Binney and Tremaine, 1987, Chap.\,4,
\S3) and two-component (e.g., Caimmi
et al., 1984; Caimmi and Secco, 1992)
systems.

The tensor and the scalar virial
theorem for the whole system read:
\begin{leftsubeqnarray}
\slabel{eq:vipa}
&& 2(E_{\rm kin})_{pq}+(E_{\rm pot})_{pq}=0~~; \\
\slabel{eq:vipb}
&& 2E_{\rm kin}+E_{\rm pot}=0~~;
\label{seq:vip}
\end{leftsubeqnarray}
which can be made explicit using
Eqs.\,(\ref{eq:pota})-(\ref{eq:potb}).
An alternative formulation can be
obtained summing Eqs.\,(\ref
{eq:vira})-(\ref{eq:virb}) over
all the subsystems.   The combination
of the related expressions yields:
\begin{leftsubeqnarray}
\slabel{eq:tica}
&& [(E_{uv})_{\rm tid}]_{pq}+[(E_{vu})_{\rm tid}]_{pq}=
[(E_{uv})_{\rm int}]_{pq}+[(E_{vu})_{\rm int}]_{pq}~~; \\
\slabel{eq:ticb}
&& (E_{uv})_{\rm tid}+(E_{vu})_{\rm tid}=
(E_{uv})_{\rm int}+(E_{vu})_{\rm int}~~; \\
\slabel{eq:ticc}
&& uv=ij,ik,jk~~;\qquad vu=ji,ki,kj~~;
\label{seq:tic}
\end{leftsubeqnarray}
according to earlier results related
to two-component systems (Caimmi and
Secco, 1992).

In the general case of subsystems with
different shapes, the tensor and scalar
virial equations, Eqs.\,(\ref{eq:vira})
and (\ref{eq:virb}), need numerical
integrations to be made explicit, except
for concentric and coaxial homogeneous
ellipsoids, one completely lying within
the other (e.g., Caimmi and Secco, 1992).
The effect of different shapes is larger
for flat density profiles, with respect
to their monotonically decreasing
counterparts.    Explicit formulae also
exist if the outer ellipsoid, in turn,
is surrounded by a similar and similarly
placed heterogeneous homeoid, where the
inner surface coincides with the boundary
of the outer ellipsoid (Caimmi and Secco,
2002).   In the more realistic situation
of homeoidally striated ellipsoids
(Roberts, 1962), a further restriction
aimed to simplify the calculations, is
that different subsystems are also similar
and similarly placed (e.g., Caimmi, 1993;
Caimmi and Marmo, 2003; Caimmi and
Valentinuzzi, 2008).

In dealing with an application of the tensor
virial theorem to a three-component system
of astrophysical interest, the choice of the
related density profiles shall be limited to
the above mentioned special cases.

\section{The model}
\label{mode}

\subsection{The hole, the vortex, and the bulge}
\label{hvsp}

Aiming to a simple application of the
tensor virial theorem, inner regions
of large galaxies shall be
idealized as three-component systems,
one completely lying within the other:
the hole, the vortex, and the
bulge, respectively.   The outer vortex
is self-gravitating, and the presence of
the inner vortex allows
mass accretion inside the last stable
orbit.   The inner bulge is enclosed
by the isopycnic (i.e. constant density)
surface with equatorial axes coinciding
with the edge of the vortex.

In absence of a satisfactory theory
of quantum gravity, the mass distribution
within the hole cannot be described,
which reduces the number of tensor
virial equations,  Eqs.\,(\ref
{eq:vira}), from nine to six.
The further restriction to an
axisymmetric gravitational potential,
implying an axisymmetric density profile,
for each subsystem, further reduces
the number of tensor virial equations
from six to four.   The final
restriction to flattened vortex
configurations, reduces the number
of tensor virial equations from
four to three.

The hole shall be modelled by a flattened homogeneous spheroid potential,
where the equatorial axis coincides with the radius of the particle inner
stable circular orbit of the related Schwartzschild black hole:
\begin{equation}
\label{eq:RH}
R_{\rm H}=3R_g~~;\qquad R_g=\frac{2GM_{\rm H}}{c^2}~~;
\end{equation}
where $G$ is the constant of gravitation, $c$ is the light velocity in
(baryonic) vacuum, $M$ is the mass, $R_g$ is the gravitational radius
(e.g., Landau and Lifchitz, 1966, Chap.\,XI, \S97), and the index, H,
denotes the hole.   The choice of a Plummer potential (Escala, 2006)
would imply a more complicated formulation (Caimmi and Valentinuzzi,
2008).   The selected value for $R_{\rm H}$, expressed by Eq.\,(\ref
{eq:RH}) and related to Schwartzschild's metrics, makes an acceptable
compromise between a corotating particle inner stable circular orbit
radius, $R_{\rm H}=(1/2)R_g$, and its counterrotating counterpart,
$R_{\rm H}=(9/2)R_g$, related to Kerr's metrics.   For the photon inner
stable circular orbit, $R_{\rm H}=(3/2)R_g$ in Schwartzschild's metrics,
and $R_{\rm H}=(1/2)R_g$ (corotating), $R_{\rm H}=2R_g$ (counterrotating),
in Kerr's metrics.   For further details refer to specialized reviews
(e.g., Bilic, 2006).

The vortex may be conceived, to a first extent, as self-gravitating
even if it is still not clear if fragmentation or turbulence are dominating
(e.g., Escala, 2006).   For simplicity, both the vortex circular
velocity, $v$, and the inner bulge velocity dispersion, $\sigma$, are
assumed constant with radius, which is only strictly valid for an isothermal
sphere (Escala, 2006).   Even within the most massive gaseous vortexes
known so far, the gravitational potential is dominated by the inner bulge.
With regard to the inner bulge, the following values are deduced from
observations (Downes and Solomon, 1998):
\begin{equation}
\label{eq:detv}
m_{\rm IN}=\frac{M_{\rm I}}{M_{\rm V}}\approx5~~;
\end{equation}
where ${M_{\rm V}}$ and ${M_{\rm I}}$ are the mass of the gaseous
vortex and stellar inner bulge, respectively.

The circular velocity at the edge of the vortex, is obtained by
the condition of centrifugal equilibrium, as:
\begin{equation}
\label{eq:v}
(v_{\rm V})^2=\frac{\zeta_{\rm V}}{m_{\rm IN}}\frac{GM_{\rm I}}{R_{\rm V}}+
\frac{\zeta_{\rm H}}{m_{\rm IH}}\frac{GM_{\rm I}}{R_{\rm V}}+
\zeta_{\rm I}\frac{GM_{\rm I}}{R_{\rm V}}~~;
\end{equation}
where $\zeta$ is a shape factor and, in particular, $\zeta=1$ for 
spherical-symmetric mass distributions and $\zeta=\pi/2$ for flattened
homogeneous spheroids.

The inner bulge velocity dispersion at the edge of the vortex
reads:
\begin{equation}
\label{eq:sigma}
(\sigma_{\rm I})^2=\frac{GM_{\rm I}}{R_{\rm V}}~~;
\end{equation}
provided the density profile declines as in an isothermal sphere.

The combination of Eqs.\,(\ref{eq:detv})-(\ref{eq:sigma}) yields:
\begin{equation}
\label{eq:vsig}
(v_{\rm V})^2=\left(\frac{\zeta_{\rm V}}{m_{\rm IN}}+
\frac{\zeta_{\rm H}}{m_{\rm IH}}+\zeta_{\rm I}\right)
(\sigma_{\rm I})^2\approx(\sigma_{\rm I})^2~~;
\end{equation}
which holds {\it a fortiori} for less massive vortexes,
$m_{\rm IN}>5$.

The validity of the vortex steady state model (the so called $\alpha$
disk) as a zeroth order approximation, has successfully been tested by
recent numerical experiments (Escala, 2006).   The vortex shall be
represented as a flattened homogeneous spheroid potential, similar and
similarly placed with respect to the hole.

Strictly speaking, the mass inside the hole volume
should be subctracted which, in the case under
discussion, amounts to:
\begin{equation}
\label{eq:DMV}
\Delta M_{\rm V}=\left(\frac{R_{\rm H}}{R_{\rm V}}\right)^3M_{\rm V}~~;
\end{equation}
and the ratio of subtracted to remaining mass reads:
\begin{equation}
\label{eq:DMM}
\frac{\Delta M_{\rm V}}{M_{\rm V}-\Delta M_{\rm V}}=\frac{(R_{\rm H}/
R_{\rm V})^3}{1-(R_{\rm H}/R_{\rm V})^3}~~;
\end{equation}
which shows that the subtracted mass, 
$\Delta M_{\rm V}$, may safely be neglected.

The inner bulge, $0\le r\le R_{\rm V}$,
shall be modelled by a homogeneous sphere potential,
and similar considerations can be made for the subctracted mass:
\begin{equation}
\label{eq:DMI}
\Delta M_{\rm I}=\left(\frac{R_{\rm H}}{R_{\rm I}}\right)^3M_{\rm I}~~;
\end{equation}
keeping in mind that $R_{\rm I}=R_{\rm V}$.

The outer bulge, $R_{\rm V}\le r\le R_{\rm B}$,
shall be modelled by a truncated isothermal sphere potential.
The related density profile (including the inner bulge) is:
\begin{leftsubeqnarray}
\slabel{eq:rhoSa}
&& \rho_{\rm B}=\rho_{\rm V}f_{\rm B}(\xi_{\rm B})~~;\qquad\xi_{\rm B}=
\frac r{R_{\rm B}}~~;\qquad (y_{\rm BV})^{-1}=\frac{R_{\rm V}}{R_{\rm B}}
~~; \\
\slabel{eq:rhoSb}
&& f_{\rm B}(\xi_{\rm B})=\cases{
1~~; & $0\le\xi_{\rm B}\le (y_{\rm BV})^{-1}~~;$ \cr
(y_{\rm BV}\xi_{\rm B})^{-2}~~; & $(y_{\rm BV})^{-1}\le\xi_{\rm B}\le1~~;$
\cr}
\label{seq:rhoS}
\end{leftsubeqnarray}
where, in general, $r_{\rm NB}=R_{\rm V}$ and $\rho_{\rm NB}=\rho_{\rm V}$
are a scaling radius and a scaling density, respectively.

In conclusion, the central part of a large galaxy shall be modelled as an
inner, flattened homogeneous spheroid (the hole), embedded within a similar
and similarly placed, flattened homogeneous spheroid (the vortex)
which, in turn, is lying within a homogeneous sphere with equal equatorial
semiaxes
(the inner bulge), surrounded by a concentric, isothermal, spherical
corona (the outer bulge).

\subsection{The tensor virial equations}
\label{tveq}

The four tensor virial equations which, among the nine expressed by
Eqs.\,(\ref{seq:vir}), are relevant to the application of interest, are:
\begin{leftsubeqnarray}
\slabel{eq:EVSa}
&& 2\left[(E_{\rm V})_{\rm kin}\right]_{\rm pp}+
\left[(E_{\rm V})_{\rm sel}\right]_{\rm pp}+
\left[(E_{\rm VH})_{\rm tid}\right]_{\rm pp}+
\left[(E_{\rm VB})_{\rm tid}\right]_{\rm pp}=0~~; \\
\slabel{eq:EVSb}
&& 2\left[(E_{\rm B})_{\rm kin}\right]_{\rm pp}+
\left[(E_{\rm B})_{\rm sel}\right]_{\rm pp}+
\left[(E_{\rm BH})_{\rm tid}\right]_{\rm pp}+
\left[(E_{\rm BV})_{\rm tid}\right]_{\rm pp}=0~~;
\label{seq:EVS}
\end{leftsubeqnarray}
where $x_3$ has been chosen as symmetry axis, which implies $p=1,3$,
and all the terms may be written explicitly, under the assumptions 
previously made.

The vortex and the bulge kinetic-energy tensors
read:
\begin{leftsubeqnarray}
\slabel{eq:EVSka}
&& 2\left[(E_{\rm V})_{\rm kin}\right]_{\rm pp}=M_{\rm V}\left\{\left(1-
\delta_{3p}\right)\left[\left(v_{\rm V}\right)_{pp}\right]^2+
\left[\left(\sigma_{\rm V}\right)_{pp}\right]^2\right\}~~; \\
\slabel{eq:EVSkb}
&& 2\left[(E_{\rm B})_{\rm kin}\right]_{\rm pp}=M_{\rm B}\left\{\left(1-
\delta_{3p}\right)\left[\left(v_{\rm B}\right)_{pp}\right]^2+
\left[\left(\sigma_{\rm B}\right)_{pp}\right]^2\right\}~~; \\
\slabel{eq:EVSkc}
&& 
v_{\rm V}\approx\sigma_{\rm B}~~;\quad p=1,3~~;
\label{seq:EVSk}
\end{leftsubeqnarray}
where $\delta$ is the Kronecker symbol, $v_{pp}$ and
$\sigma_{pp}$ are the circular velocity and the random velocity
components along the corresponding axis, respectively.

In the special case of homeoidally striated ellipsoids
(Roberts, 1962), the self-potential energy tensors
read (e.g., Caimmi, 1993):
\begin{leftsubeqnarray}
\slabel{eq:EVSsa}
&& \left[(E_{\rm X})_{\rm sel}\right]_{\rm pp}=-\left(\nu_{\rm X}\right)_{\rm
sel}\frac{GM_{\rm X}^2}{R_{\rm X}}(B_{\rm X})_p~~;\qquad {\rm X=V,B}~~; \\
\slabel{eq:EVSsb}
&& B_p=\epsilon_{p2}\epsilon_{p3}\int_0^{+\infty}(1+s)^{-3/2}(1+\epsilon_
{p\ell}^2s)^{-1/2}(1+\epsilon_{pr}^2s)^{-1/2}\diff s~~; \\
\slabel{eq:EVSsc}
&& \epsilon_{pq}=\frac{a_p}{a_q}~~;\qquad p=1,2,3~~;\qquad q=1,2,3~~;\qquad
p\ne\ell\ne r~~;
\label{seq:EVSs}
\end{leftsubeqnarray}
where $R$ is
the major equatorial semiaxis, $\left(\nu_{\rm X}\right)_
{\rm sel}$ are profile factors, $B_p$ shape factors
which, for axisymmetric configurations, may be analytically
expressed (e.g., Chandrasekhar, 1969, Chap.\,3, \S21;
Caimmi, 1991, 1995), and $\epsilon_{pq}$ axis ratios.
For further details refer to Appendix\,\ref{a:sfa}.

The vortex self potential energy profile factor reads
(Caimmi and Secco, 1992):
\begin{equation}
\label{eq:nuVs}
\left(\nu_{\rm V}\right)_{\rm sel}=\frac3{10}~~;
\end{equation}
and the shape factors are expressed in 
Appendix\,\ref{a:sfa}.

The bulge self potential energy profile factor reads
(Caimmi and Secco, 2002)%
\footnote{The related formula in the parent paper,
Eq.\,(51) therein, is different due to printing
errors.}:
\begin{equation}
\label{eq:nuSs}
\left(\nu_{\rm B}\right)_{\rm sel}=\frac3{10}\frac{15(y_{\rm BV})^2-14
y_{\rm BV}-10y_{\rm BV}\ln(y_{\rm BV})}{(3y_{\rm BV}-2)^2}~~;
\end{equation}
related to the density profile, expressed
by Eq.\,(\ref{seq:rhoS}).

In the general case of triaxial configurations,
the tidal potential-energy tensors depend on the
density profiles and on the shape of the ellipsoids.
The isopycnic surface of the outer ellipsoid which
is still embedding the inner one, is tangent to the
latter boundary at a top axis,
$(a_i)_t$.  For oblate-like configurations,
the inner more flattened than the outer, $(a_i)_t=
(a_i)_1$.

The mass of the outer ellipsoid surrounded by the
isothermal homeoid, is (Caimmi and Secco, 2002)%
\footnote{With regard to the profile factor, the
related formula in the parent paper,
Eq.\,(49) therein, is different due to printing
errors.   In addition, $\xi_k^2$ has to be replaced
by $\xi_k^3$ in Eq.\,(50) therein.}:
\begin{leftsubeqnarray}
\slabel{eq:Smasa}
&& M_{\rm B}=\left(\nu_{\rm B}\right)_{\rm mas}(M_{\rm B})_0~~; \\
\slabel{eq:Smasb}
&& (M_{\rm B})_0=\frac{4\pi}3(\rho_{\rm B})_0(a_{\rm B})_1(a_{\rm B})_2
(a_{\rm B})_3~~; \\
\slabel{eq:Smasc}
&& \left(\nu_{\rm B}\right)_{\rm mas}=(y_{\rm BV})^{-3}[1+3(y_{\rm BV}-1)]
~~; \\
\slabel{eq:Smasd}
&& y_{\rm BV}=(y_{\rm BV})_t=\frac{(a_{\rm B})_t}{(a_{\rm V})_t}~~;
\label{seq:Smas}
\end{leftsubeqnarray}
where $(\rho_{\rm B})_0$ is the central density,
$(M_{\rm B})_0$ the mass of a homogeneous ellipsoid
with same boundary and density equal to the central
density, and the major axis of the inner bulge
and the vortex coincide.   Accordingly,
Eq.\,(\ref{eq:Smasa}) takes the equivalent form:
\begin{equation}
\label{eq:SSmas}
M_{\rm B}=M_{\rm I}[1+3(y_{\rm BV}-1)]~~;
\end{equation}
where $M_{\rm I}$ is the mass of the inner bulge.

In the special case of two concentric and coaxial
homogeneous ellipsoids, the more flattened one
completely lying within the other with a tangential
point at the top major equatorial semiaxis, $R_{\rm
X}$, surrounded by a isothermal (X=V) or homogeneous
+ isothermal (X=H) homeoid where the
inner surface coincides with the boundary of the
outer ellipsoid, the tidal potential-energy tensors
read (Caimmi and Secco, 2001, 2002):
\begin{leftsubeqnarray}
\slabel{eq:EXYta}
&& \left[(E_{\rm XY})_{\rm tid}\right]_{pq}=-\delta_{pq}\frac{GM_{\rm X}^2}
{R_{\rm X}}\left(\nu_{\rm XY}\right)_{\rm tid}\frac{[(y_{\rm
YX})_1]^2}{[(y_{\rm YX})_p]^2}(B_{\rm Y})_p~~; \\
\slabel{eq:EXYtb}
&& \left[(E_{\rm YX})_{\rm tid}\right]_{pq}=-\delta_{pq}\frac{GM_{\rm X}^2}
{R_{\rm X}}\left(\nu_{\rm XY}\right)_{\rm tid}\left\{\frac52
[(y_{\rm YX})_1]^2F_{\rm Y}(y_{\rm YX})+(\Phi_{\rm YX})_p\right\}(B_
{\rm Y})_p;\qquad \\
\slabel{eq:EXYtc}
&& \left(\nu_{\rm XY}\right)_{\rm tid}=\frac3{10}\frac{m_{\rm YX}}
{[(y_{\rm YX})_1]^3}\frac1{\left(\nu_{\rm Y}\right)_{\rm mas}}~;\quad
m_{\rm YX}=\frac{M_{\rm Y}}{M_{\rm X}}~;\quad(y_{\rm YX})_p=\frac
{(a_{\rm Y})_p}{(a_{\rm X})_p}>1~; \\
\slabel{eq:EXYtd}
&& \frac{(\epsilon_{\rm X})_{pq}}{(\epsilon_{\rm Y})_{pq}}=
\frac{(y_{\rm YX})_q}{(y_{\rm YX})_p}~;\quad{\rm X=H,V}~;\quad{\rm Y=B}~;
\quad p=1,2,3~;\quad q=1,2,3~; \\
\slabel{eq:EXYte}
&& F_{\rm B}(y_{\rm BV})=2\frac{\ln(y_{\rm BV})}{(y_{\rm BV})^2}~~; \qquad
F_{\rm B}(y_{\rm BH})=\frac1{(y_{\rm BV})^2}-\frac1{(y_{\rm BH})^2}+
2\frac{\ln(y_{\rm BV})}{(y_{\rm BV})^2}~~; \\
\slabel{eq:EXYtg}
&& (\Phi_{\rm YX})_p=\frac52-\frac32\left[\frac{(y_{\rm YX})_1}
{(y_{\rm YX})_p}\right]^2-\frac12\sum_{r=1}^3\left\{\left[\frac
{(y_{\rm YX})_1}{(y_{\rm YX})_r}\right]^2-\left[\frac{(y_{\rm YX})_1}
{(y_{\rm YX})_p}\right]^2\right\}\frac{(B_{\rm Y})_{pr}}{(B_{\rm Y})_p}~~; \\
\slabel{eq:EXYth}
&& B_{pr}=\epsilon_{p2}\epsilon_{p3}\int_0^{+\infty}(1+s)^{-3/2}(1+\epsilon_
{p\ell}^2s)^{-1/2}(1+\epsilon_{pr}^2s)^{-3/2}\diff s~~; \\
\slabel{eq:EXYti}
&& p=1,2,3~~;\qquad \ell=1,2,3~~;\qquad r=1,2,3~~;\qquad p\ne\ell\ne r~~;
\label{seq:EXYt}
\end{leftsubeqnarray}
where $\nu_{\rm tid}$ are profile factors and $B_{pr}$
are shape factors
which, for axisymmetric configurations, may be analytically
expressed (e.g., Chandrasekhar, 1969, Chap.\,3, \S21;
Caimmi, 1995).
For further details refer to Appendix\,\ref{a:sfa}.

In the special case of two concentric and coaxial
homogeneous ellipsoids, one completely lying within
the other (e.g., Caimmi and Secco, 1992), the tidal
potential-energy tensors read:
\begin{leftsubeqnarray}
\slabel{eq:EHVa}
&& \left[(E_{\rm HV})_{\rm tid}\right]_{pq}=-\delta_{pq}\frac{GM_{\rm H}^2}
{R_{\rm H}}\left(\nu_{\rm HV}\right)_{\rm tid}\frac{[(y_{\rm VH})_1]^2}
{[(y_{\rm VH})_p]^2}(B_{\rm V})_p~~; \\
\slabel{eq:EHVb}
&& \left[(E_{\rm VH})_{\rm tid}\right]_{pq}=-\delta_{pq}
\frac{GM_{\rm H}^2}{R_{\rm H}}\left(\nu_{\rm HV}\right)_{\rm tid}\left[
F_{\rm V}(y_{\rm VH})
+(\Phi_{\rm VH})_p\right](B_{\rm V})_p;\qquad \\
\slabel{eq:EHVc}
&& \left(\nu_{\rm HV}\right)_{\rm tid}=\frac3{10}\frac{m_{\rm VH}}
{[(y_{\rm VH})_1]^3}\frac1{\left(\nu_{\rm V}\right)_{\rm mas}}~;\quad
\left(\nu_{\rm H}\right)_{\rm mas}=\left(\nu_{\rm V}\right)_{\rm mas}=1~; \\
\slabel{eq:EHVd}
&& F_{\rm V}(y_{\rm VH})=\frac52
\left[\frac{(y_{\rm VH})_1}{(y_{\rm VH})_p}\right]^2\left\{
[(y_{\rm VH})_p]^2-\left[\frac{(y_{\rm VH})_p}{(y_{\rm VH})_t}\right]^2
\right\}~~;
\label{seq:EHV}
\end{leftsubeqnarray}
which may be conceived as a three-component system,
as discussed above, where the surrounding homeoid
is homogeneous instead of isothermal or homogeneous
+ isothermal.

In the special case under discussion, the shape factors
take the expression:
\begin{leftsubeqnarray}
\slabel{eq:phi1a}
&& (\Phi_{\rm YX})_1=(\Phi_{\rm YX})_2=1+\frac3{10}(1-\epsilon^2)~~; \\
\slabel{eq:phi1b}
&& (\Phi_{\rm YX})_3=1+\frac9{10}(1-\epsilon^2)~~; \\
\slabel{eq:phi1c}
&& (\Phi_{\rm VH})_1=(\Phi_{\rm VH})_2=(\Phi_{\rm VH})_3=1~~; \\
\slabel{eq:phi1d}
&& \epsilon_{\rm H}=\epsilon_{\rm V}=\epsilon~~;\qquad\epsilon_{\rm B}=1~~;
\\
\slabel{eq:phi1e}
&& \frac{(a_{\rm V})_p}{(a_{\rm V})_r}=\frac{(a_{\rm H})_p}{(a_{\rm H})_r}~~;
\qquad (y_{\rm VH})_p=(y_{\rm VH})_r=y_{\rm VH}~~;
\label{seq:phi1}
\end{leftsubeqnarray}
and the four tensor virial equations of interest,
Eqs.\,(\ref{seq:EVS}), due to (\ref{seq:EVSk})-(\ref
{seq:phi1}), may be written under the explicit form:
\begin{leftsubeqnarray}
\slabel{eq:vsa}
&& \left[(v_{\rm V})_{qq}\right]^2+\left[(\sigma_{\rm V})_{qq}\right]^2=
\frac3{10}\frac{GM_{\rm V}}{R_{\rm V}}\left\{\frac\alpha\epsilon\left[1+
\frac1{m_{\rm VH}(y_{\rm VH})^2}\left(\frac52(y_{\rm VH})^2-\frac32\right)
\right]\right. \nonumber \\
&& \phantom{(v_{\rm V})_{qq}^2+=}
\left.+\frac23\frac{m_{\rm BV}}{3y_{\rm BV}-2}\left[\frac
{(y_{\rm BV})_1}{(y_{\rm BV})_q}\right]^2\left[\frac{(y_{\rm BV})_t}
{(y_{\rm BV})_1}\right]^3\right\}~;\qquad q=1,2~~; \\
\slabel{eq:vsb}
&& \left[(\sigma_{\rm V})_{33}\right]^2=
\frac3{10}\frac{GM_{\rm V}}{R_{\rm V}}\left\{\epsilon\gamma\left[1+
\frac1{m_{\rm VH}(y_{\rm VH})^2}\left(\frac52(y_{\rm VH})^2-\frac32\right)
\right]\right. \nonumber \\
&& \phantom{\left[(\sigma_{\rm V})_{qq}
\right]^2=} \left.+\frac23\frac{m_{\rm BV}}{3(y_{\rm BV})_t-2}\left[\frac
{(y_{\rm BV})_1}{(y_{\rm BV})_3}\right]^2\left[\frac{(y_{\rm BV})_t}
{(y_{\rm BV})_1}\right]^3\right\}~; \\
\slabel{eq:vsc}
&& \left[(v_{\rm B})_{qq}\right]^2+\left[(\sigma_{\rm B})_{qq}\right]^2=
\frac15\frac{GM_{\rm B}}{R_{\rm B}}\left\{\frac{15(y_{\rm BV})^2-
14y_{\rm BV}-10y_{\rm BV}\ln(y_{\rm BV})}{(3y_{\rm BV}-2)^2}
\right. \nonumber \\
&& +\frac1{m_{\rm BH}(y_{\rm VH})^2}\left[\frac{(y_{\rm BH})_t}
{(y_{\rm BH})_1}\right]^2\frac{y_{\rm BV}}{3y_{\rm BV}-2}
\left\{\frac52\left[\frac{(y_{\rm BH})_1}{(y_{\rm BH})_t}\right]^2
\right. \nonumber \\
&& \left.\times\left[(y_{\rm VH})^2-1+2(y_{\rm VH})^2\ln(y_{\rm BV})\right]
+1+\frac3{10}\left(1-\epsilon^2\right)\right\}
+\frac1{m_{\rm BV}}\left[\frac{(y_{\rm BV})_t}{(y_{\rm BV})_1}\right]^2
\nonumber \\
&& \left.\times
\frac{y_{\rm BV}}{3y_{\rm BV}-2}\left[5
\left[\frac{(y_{\rm BV})_1}{(y_{\rm BV})_t}\right]^2
\ln(y_{\rm BV})+1+\frac3{10}
\left(1-\epsilon^2\right)\right]\right\}~;\quad q=1,2~; \\
\slabel{eq:vsd}
&& \left[(\sigma_{\rm B})_{33}\right]^2=
\frac15\frac{GM_{\rm B}}{R_{\rm B}}\left\{\frac{15(y_{\rm BV})^2-
14y_{\rm BV}-10y_{\rm BV}\ln(y_{\rm BV})}{(3y_{\rm BV}-2)^2}
\right. \nonumber \\
&& +\frac1{m_{\rm BH}(y_{\rm VH})^2}
\left[\frac{(y_{\rm BH})_t}{(y_{\rm BH})_1}\right]^2
\frac{y_{\rm BV}}{3y_{\rm BV}-2}
\left\{\frac52\left[\frac{(y_{\rm BH})_1}{(y_{\rm BH})_t}\right]^2
\right.\nonumber \\
&& \left.\times
\left[(y_{\rm VH})^2-1+2(y_{\rm VH})^2\ln(y_{\rm BV})\right]+
1+\frac9{10}\left(1-\epsilon^2\right)\right\}
+\frac1{m_{\rm BV}}
\left[\frac{(y_{\rm BV})_t}{(y_{\rm BV})_1}\right]^2 \nonumber \\ 
&& \left.\times
\frac{y_{\rm BV}}{3y_{\rm BV}-2}\left[5
\left[\frac{(y_{\rm BV})_1}{(y_{\rm BV})_t}\right]^2
\ln(y_{\rm BV})+1+\frac9{10}
\left(1-\epsilon^2\right)\right]\right\}~~;
\label{seq:vs}
\end{leftsubeqnarray}
where, in the special case of flattened configurations,
$\epsilon=0$, the shape factors reduce to $\alpha/
\epsilon=\pi/2$ and $\epsilon\gamma=0$, see
Appendix\,\ref{a:sfa}, which yields $(\sigma_
{\rm V})_{33}=0$, as expected.   The validity of
Eqs.\,(\ref{eq:vsc}) and (\ref{eq:vsd}) implies an
anisotropic bulge stress tensor, $\left[(v_{\rm B})_
{qq}\right]^2+\left[(\sigma_{\rm B})_{qq}\right]^2<
\left[(\sigma_{\rm B})_{33}\right]^2$, which is due
to the presence of more flattened, inner subsystems.
In fact, an anisotropic stress tensor is necessary
to maintain a spherical shape in the case under
discussion.

The tensor virial equations,
Eqs.\,(\ref{eq:vsa})-(\ref{eq:vsd}), make a system
of four equations in thirteen unknowns: two rotation
velocity components, $(v_{\rm V})_{11}=(v_{\rm V})_
{22}$ and $(v_{\rm B})_{11}=(v_{\rm B})_{22}$; four
peculiar velocity components, $(\sigma_{\rm V})_{11}
=(\sigma_{\rm V})_{22}$, $(\sigma_{\rm V})_{33}$,
$(\sigma_{\rm B})_{11}=(\sigma_{\rm B})_{22}$, and
$(\sigma_{\rm B})_{33}$; three masses, $M_{\rm H}$,
$M_{\rm V}$, and $M_{\rm B}$; three major equatorial
semiaxes, $R_{\rm H}$, $R_{\rm V}$, and $R_{\rm B}$;
and one axis ratio, $\epsilon=\epsilon_{\rm H}=
\epsilon_{\rm V}$.   The hole mass, $M_{\rm H}$,
and the model hole major semiaxis, $R_{\rm H}$, are
connected via Eq.\,(\ref{eq:RH}), which reduces
the number of unknowns to twelve.   Further
reduction can be made using observational constraints
and/or additional assumptions.

Rotation and peculiar velocity components
could be deduced from observations,
together with the vortex and
bulge major semiaxes.   The remaining four
unknowns would be three masses and one
axis ratio.   In the flat limit $(\epsilon
\to0)$ Eq.\,(\ref{eq:vsb}) reduces to an
indeterminate form, $0=0$, leaving three
equations in three unknowns, $M_{\rm H}$,
$M_{\rm V}$, and $M_{\rm B}$.   Of course,
any other alternative might be exploited,
yielding a system of four (or three)
equations in four (or three) unknowns.

\subsection{Application}
\label{app}

According to an earlier model of massive
vortexes (Escala, 2006), the hole
and the vortex mass shall be taken
as $M_{\rm H}=10^7{\rm m}_\odot$ and
$M_{\rm V}=5\cdot10^7{\rm m}_\odot$,
respectively.   A flat disk $(\epsilon_
{\rm H}=\epsilon_{\rm V}=0)$ makes an
acceptable approximation to an assumed
major equatorial semiaxis,
$R_{\rm V}=125{\rm pc}$, and thickness,
$Z_{\rm V}=4{\rm pc}$, (Escala, 2006)
implying $\epsilon_{\rm V}=Z_{\rm V}/
R_{\rm V}=0.032$.
The particularization of Eq.\,(\ref
{eq:RH}) to the
selected hole mass yields: $R_{\rm H}=
2.87\cdot10^{-6}{\rm pc}$.

The related parameters appearing in
Eqs.\,(\ref{eq:vsa})-(\ref{eq:vsd})
are:
\begin{equation}
\label{eq:pam}
m_{\rm VH}=5~~;\qquad y_{\rm VH}=4.35\cdot10^7~~;\qquad\epsilon=0~~;
\end{equation}
and the substitution into the above
mentioned equations, keeping in mind
that Eq.\,(\ref{eq:vsb}) cannot be
used for flat configurations, yields:
\begin{leftsubeqnarray}
\slabel{eq:vspa}
&& \left[(v_{\rm V})_{qq}\right]^2+\left[(\sigma_{\rm V})_{qq}\right]^2=
\frac15\frac{GM_{\rm V}}{R_{\rm V}}\left\{\frac{3\pi}4\left(1+\frac52\frac1
{m_{\rm VH}}\right)+\frac{m_{\rm BV}}{3y_{\rm BV}-2}
\right\}~~;
\\
\slabel{eq:vspb}
&& \left[(v_{\rm B})_{qq}\right]^2+\left[(\sigma_{\rm B})_{qq}\right]^2=
\frac15\frac{GM_{\rm B}}{R_{\rm B}}\left\{\frac{15(y_{\rm BV})^2-
14y_{\rm BV}-10y_{\rm BV}\ln(y_{\rm BV})}{(3y_{\rm BV}-2)^2}
\right. \nonumber \\
&& 
\left.+\frac1{m_{\rm BH}}\frac{y_{\rm BV}}{3y_{\rm BV}-2}\frac52\left[1+2
\ln(y_{\rm BV})\right]
+\frac1{m_{\rm BV}}\frac{y_{\rm BV}}{3y_{\rm BV}-2}\left[5\ln(y_{\rm BV})+
\frac{13}{10}\right]\right\}~; \nonumber \\
&&  \\
\slabel{eq:vspc}
&& \left[(\sigma_{\rm B})_{33}\right]^2=
\frac15\frac{GM_{\rm B}}{R_{\rm B}}\left\{\frac{15(y_{\rm BV})^2-
14y_{\rm BV}-10y_{\rm BV}\ln(y_{\rm BV})}{(3y_{\rm BV}-2)^2}
\right. \nonumber \\
&& 
\left.+\frac1{m_{\rm BH}}\frac{y_{\rm BV}}{3y_{\rm BV}-2}\frac52\left[1+2
\ln(y_{\rm BV})\right]
+\frac1{m_{\rm BV}}\frac{y_{\rm BV}}{3y_{\rm BV}-2}\left[5\ln(y_{\rm BV})+
\frac{19}{10}\right]\right\}~~; \nonumber \\
&&
\label{seq:vsp}
\end{leftsubeqnarray}
where the terms containing negative powers of
$y_{\rm VH}\gg1$ and $y_{\rm BH}>y_{\rm VH}$,
have been neglected.

In the current model, the vortex is described
as a homogeneous, self-gravitating, flattened spheroid,
which implies the relations:
\begin{leftsubeqnarray}
\slabel{eq:OvNa}
&& \frac12I_{\rm V}\Omega_{\rm V}^2=\frac12M_{\rm V}\overline{[(v_{\rm V})^
2]}=2\frac{(E_{\rm V})_{\rm kin}}{M_{\rm V}}~~;\qquad(\sigma_{\rm V})_{qq}
\ll(v_{\rm V})_{qq}~~; \\
\slabel{eq:OvNb}
&& I_{\rm V}\Omega_{\rm V}^2=\frac25M_{\rm V}(R_{\rm V})^2\Omega_{\rm V}^2=
\frac25M_{\rm V}[v_{\rm V}(R_{\rm V})]^2~~;
\label{seq:OvN}
\end{leftsubeqnarray}
where $\Omega_{\rm V}$ is the (solid-body) angular
velocity, $\overline{[(v_{\rm V})^2]}$ the mean
square rotation velocity, and $v_{\rm V}(R_{\rm V})=
\{2.5\overline{[(v_{\rm V})^2]}\}^{1/2}$ the rotation
velocity at the edge.   Strictly speaking, the
presence of the hole implies differential rotation
towards Keplerian velocities, but this effect can
be neglected beyond the hole sphere of influence
(e.g., Ferrarese and Ford, 2005).
For the selected reference
case, Eqs.\,(\ref{seq:vsp}) yield $[(v_{\rm V})_
{qq}]^2/[(\sigma_{\rm V})_{33}]^2\approx0.4$ or
$\{[v_{\rm V}(R_{\rm V})]_{qq}\}^2/[(\sigma_{\rm V})_{33}]^2
\approx1$, which is consistent with Eq.\,(\ref{eq:vsig}),
as expected for massive inner bulges (e.g., Escala, 2006).

In the limit of null vortex and hole mass,
Eq.\,(\ref{eq:vspc}) reduces to:
\begin{equation}
\label{eq:sigB0}
\left[(\sigma_{\rm B0})_{33}\right]^2=
\frac15\frac{GM_{\rm B}}{R_{\rm B}}\frac{15(y_{\rm BV})^2-
14y_{\rm BV}-10y_{\rm BV}\ln(y_{\rm BV})}{(3y_{\rm BV}-2)^2}~~;
\end{equation}
and the combination of Eqs.\,(\ref{eq:vspc}) and
(\ref{eq:sigB0}) yields:
\begin{leftsubeqnarray}
\slabel{eq:MHsBa}
&& \frac1{m_{\rm BH}}=\frac{M_{\rm H}}{M_{\rm B}}=\frac{R_{\rm B}\left[
(\sigma_{\rm B})_{33}\right]^2}{GM_{\rm B}}\left\{1-\left[\frac{(\sigma_
{\rm B0})_{33}}{(\sigma_{\rm B})_{33}}\right]^2\right\}\zeta(m_{\rm BH},
y_{\rm BV})~~; \\
\slabel{eq:MHsBb}
&& \zeta(m_{\rm BH},y_{\rm BV})=\frac{3y_{\rm BV}-2}{y_{\rm BV}}\left[\frac
12+(1+m_{\rm VH})\ln(y_{\rm BV})+\frac{19}{50}m_{\rm VH}\right]^{-1};
\label{seq:MHsB}
\end{leftsubeqnarray}
where in the range of interest, $1\le y_{\rm BV}\le10^4$,
$0\le m_{\rm VH}\le5$, the contribution of the factor,
$\zeta(m_{\rm BH},y_{\rm BV})$, scales from about 0.05
to 2, related to $(m_{\rm VH},y_{\rm BV})=(5,10^4), (0,1)$,
respectively.   A selected reference case, $(m_{\rm VH},
y_{\rm BV})=(5,14)$, yielding $M_{\rm B}=10^{10}{\rm m}_
\odot$ for an assumed $M_{\rm H}=10^7{\rm m}_\odot$,
corresponds to $\zeta(5,14)=0.156690$.   In the limit
of an infinitely extended bulge, $y_{\rm BV}\to+\infty$,
$\zeta(m_{\rm BH},y_{\rm BV})\to0$, due to $M_{\rm B}\to
+\infty$, for the isothermal sphere.   Then the fractional
mass, $m_{\rm BH}$, to a first extent, may be considered
as independent of
the vortex mass and bulge to vortex major
semiaxis ratio.

The mere existence of a fundamental plane (Djorgovski
and Davis, 1987; Dressler et al., 1987) indicates that
structural properties in spheroids (elliptical galaxies
and spiral bulges) span a narrow range, suggesting some
self-regulating mechanism must be at work during formation
and evolution.   In particular, projected light profiles
exhibit a large degree of homogeneity and may well be
fitted by the $r^{1/4}$ de Vaucouleurs law.   Accordingly,
a narrow range may safely be expected also for the factor,
$1-\left[(\sigma_{\rm B0})_{33}/(\sigma_{\rm B})_{33}
\right]^2$, appearing in Eq.\,(\ref{seq:MHsB}), which makes
the hole mass, $M_{\rm H}$, depend only on the product,
$R_{\rm B}\left[(\sigma_{\rm B})_{33}\right]^2$,
to a first extent.

If, in addition, the bulge density profile is close to
an isothermal sphere, then the central velocity dispersion
projected on the line of sight, $\sigma_0$, is
close to the rms velocity dispersion averaged on the mass,
$(\sigma_{\rm B})_{33}$.   Then the hole mass depends only
on the product, $R_e\sigma_0^2$,
to a first extent.   Using the Faber-Jackson relation,
$R_e\propto\sigma_0^2$,
the standard $M_{\rm H}$-$\sigma_0$
form $(M_{\rm H}\propto\sigma_0^4)$
is obtained (Escala, 2006).

A different analysis can be performed fixing the bulge
mass, $M_{\rm B}$, or the bulge to vortex mass
ratio, $m_{\rm BV}$, and considering the velocity
ratio, $\eta=(\sigma_{\rm B})_{33}/\{
[(v_{\rm V})_{qq}]^2+[(\sigma_{\rm V})_{qq}]^2\}^{1/2}$,
as a function of the bulge to vortex major axis
ratio, $y_{\rm BV}$.   The related explicit expression
is obtained by the combination of Eqs.\,(\ref{seq:phi1}),
(\ref{eq:vsa}), (\ref{eq:vsd}), particularized to flat
inner subsystems, which represents
isofractional mass (i.e. fixed $m_{\rm BV}$, $m_
{\rm BH}$, $m_{\rm VH}$) curves in the $({\sf O}
y_{\rm BV}\eta)$ plane.

\begin{figure*}[t]
\begin{center}
\includegraphics[scale=0.8]{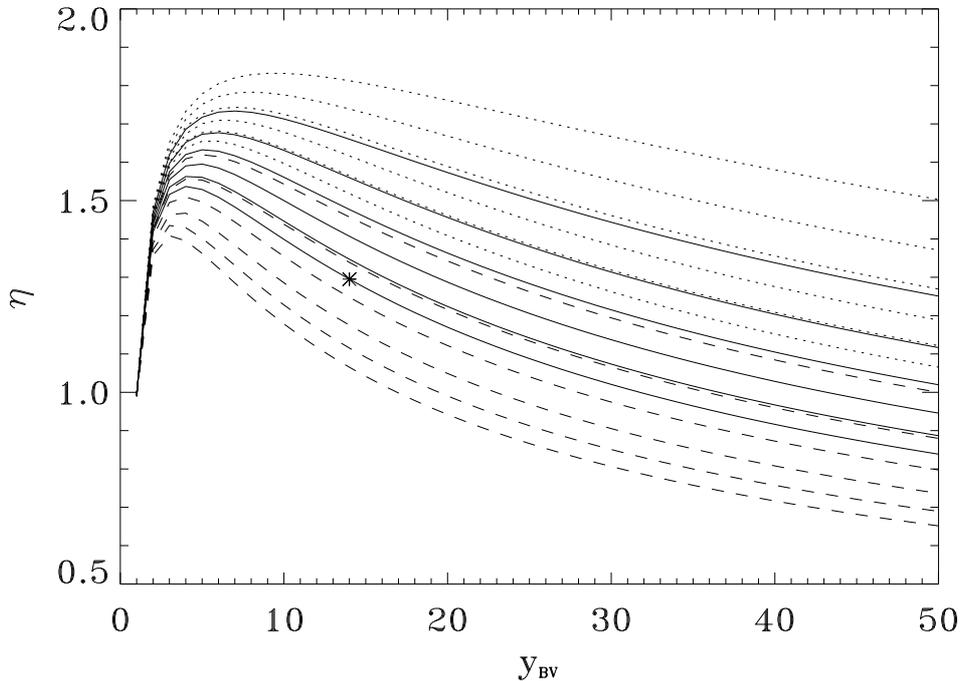}
\caption{Isofractional mass curves in the $({\sf O}y_
{\rm BV}\eta)$ plane for $m_{\rm BH}=427$
(full curves), which represents the preferred value of
a recent interpolation of the empirical
$M_{\rm H}$-$M_{\rm B}$ relation (Marconi and Hunt,
2003).    Counterparts corresponding to an estimated upper
($m_{\rm BH}=229$, dashed curves) and lower ($m_{\rm BH}=
795$, dotted curves) limit to the $M_{\rm H}$-$M_{\rm B}$
relation (Marconi and Hunt, 2003), are also shown.
Different curves
of each class (from top to bottom) correspond to
$m_{\rm VH}=0,1,2,3,4,5$, respectively.    The
reference configuration, $(m_{\rm BH},m_{\rm VH},
y_{\rm BV})=(427,200,14)$, is represented by an
asterisk.}
\label{f:isfm}
\end{center}    
\end{figure*}

The fractional masses, by definition, are related
as $m_{\rm BV}=m_{\rm BH}/m_{\rm VH}$, where $0\le
m_{\rm VH}\le5$ in the cases of interest.   On the
other hand, a correlation between bulge and hole
mass, yielding $229<m_{\rm BH}<795$, with a preferred
value, $m_{\rm BH}\approx427$, has been
deduced from recent observations (Marconi and Hunt,
2003).   The assumption of constant $m_{\rm BH}$
implies isofractional mass curves
in the $({\sf O}y_{\rm BV}\eta)$ plane depend on a
single fractional mass, $m_{\rm VH}$.  
Different cases in the range of interest,
for $m_{\rm BH}=427$,
are shown by full curves plotted in Fig.\,\ref{f:isfm}.
Counterparts corresponding to an estimated upper
($m_{\rm BH}=229$) and lower ($m_{\rm BH}=795$)
limit to the $M_{\rm H}$-$M_{\rm B}$
relation (Marconi and Hunt, 2003), are shown by
dashed and dotted curves, respectively.
Different curves
of each class (from top to bottom) correspond to
$m_{\rm VH}=0,1,2,3,4,5$, respectively.    The
reference configuration, $(m_{\rm BH},m_{\rm VH},
y_{\rm BV})=(427,5,14)$, is represented by an
asterisk.

An inspection of Fig.\,\ref{f:isfm} discloses
the following features.
\begin{description}
\item[\rm{(i)}] All curves exhibit an extremum point
of maximum.   Lower $m_{\rm BH}$ and/or $m_{\rm VH}$
correspond to lower maximum coordinates, and vice versa.
\item[\rm{(ii)}] All curves are close, one with respect
to the other, when they are rising, on the left of the
maximum.
\item[\rm{(iii)}] All curves are parallel, one with
respect to the other, when they are declining, on the
right of the maximum.
\item[\rm{(iv)}] Curves related to different fractional
masses, $m_{\rm BH}$ and $m_{\rm VH}$, could be nearly
coincident for a convenient choice of values.   This is
the case for $(m_{\rm BH},m_{\rm VH})=(427,1)$
and $(795,4)$, plotted in Fig.\,\ref{f:isfm} as full and dotted
lines, respectively.
\end{description}

\subsection{Special cases}
\label{spec}

Aiming to see the model at work, two special
cases shall be considered, where vortex
rotation curves have been deduced from
observations: NGC 4374 (M84) and NGC 4486
(M87).   Parameter values of interest are
listed in Tab.\,\ref{t:spec}.
\begin{table}
\begin{tabular}{lllll}
\hline
\hline
\multicolumn{1}{l}{NGC/M} &
\multicolumn{1}{c}{4374/84} &
\multicolumn{1}{c}{}
&
\multicolumn{1}{c}{4486/87} &
\multicolumn{1}{c}{} \\
\hline
$R_e$/kpc                            & 6.15$\mp$1.88          & (1) & 7.96$\mp$3.23 & (1) \\
$\sigma_e$/(km\,s$^{-1})$            & 278$\mp$14             & (1) & 298$\mp$15    & (1) \\
$R_{\rm V}^\prime$/pc                & 164$\mp$50             & (2) & 35$\mp$14     & (3) \\
$V_{\rm V}\sin i/({\rm km\,s}^{-1}$) & 250$\mp$150            & (2) & 500$\mp$100   & (3) \\
$i$/deg                              & 80$\mp$5               & (2) & 51$\mp$5      & (3) \\
$M_{\rm H}/{\rm M}_{10}$             & $0.15_{+0.11}^{-0.06}$ & (2) & 0.32$\mp$0.09 & (3) \\
\hline\hline
\end{tabular}
\caption{Parameter values related to NGC 4374
(M84) and NGC 4486 (M87).   Captions: $R_e$ -
bulge effective (half-light) radius; $\sigma_e$
- luminosity-weighted second moment of the
line-of-sight velocity dispersion within the
effective radius; $R_{\rm V}^\prime$ - vortex radius
at the end of the rotation velocity profile;
$V_{\rm V}$ - vortex rotation velocity weighted
on the mass, $M_{\rm V}V_{\rm V}^2=\int_{S_{\rm
V}}\rho_{\rm V}v_{\rm rot}^2\diff S_{\rm V}$;
$i$ - inclination angle; $M_{\rm H}$ - hole
mass; ${\rm M}_{10}=10^{10}{\rm m}_\odot$.
See text for details.   The velocity conversion
factors from [cm\,g\,s] to [kpc\,M$_{10}$\,Gyr]
system of measure and vice versa are:
1\,kpc\,Gyr$^{-1}$=0.978\,461\,942\,118\,946\,6 km\,s$^{-1}$ and
1\,km\,s$^{-1}$=1.022\,012\,156\,992\,443 kpc\,Gyr$^{-1}$,
respectively.   References: (1) Cappellari et al.
(2006); (2) Bower et al. (1998); (3) Macchetto et
al. (1997).}
\label{t:spec}
\end{table}
Bulge data and related uncertainties are
taken from a recent investigation
(Cappellari et al., 2006), and errors
are propagated using a standard quadratic
formula.   Vortex data are taken from
different sources for NGC 4374 (Bower
et al., 1998) and NGC 4486 (Macchetto
et al., 1997).   Rotation velocities
weighted on the mass are roughly estimated
as the mean of the maximum and the
plateau value of the rotation curve,
with no correction for inclination
effects, whose contribution is thought
to be negligible in the framework of
the above approximation.   The related
uncertainty is assumed to be equal to the
deviation from the mean.   Radii correspond
to the end of measured rotation curves.   The
related uncertainty is assumed to be equal
(in percentage) to its bulge counterpart,
$\sigma_{\rm R_V^\prime}/R_{\rm V}^\prime=
\sigma_{\rm R_B}/R_{\rm B}$.

\begin{figure*}[t]
\begin{center}
\includegraphics[scale=0.8]{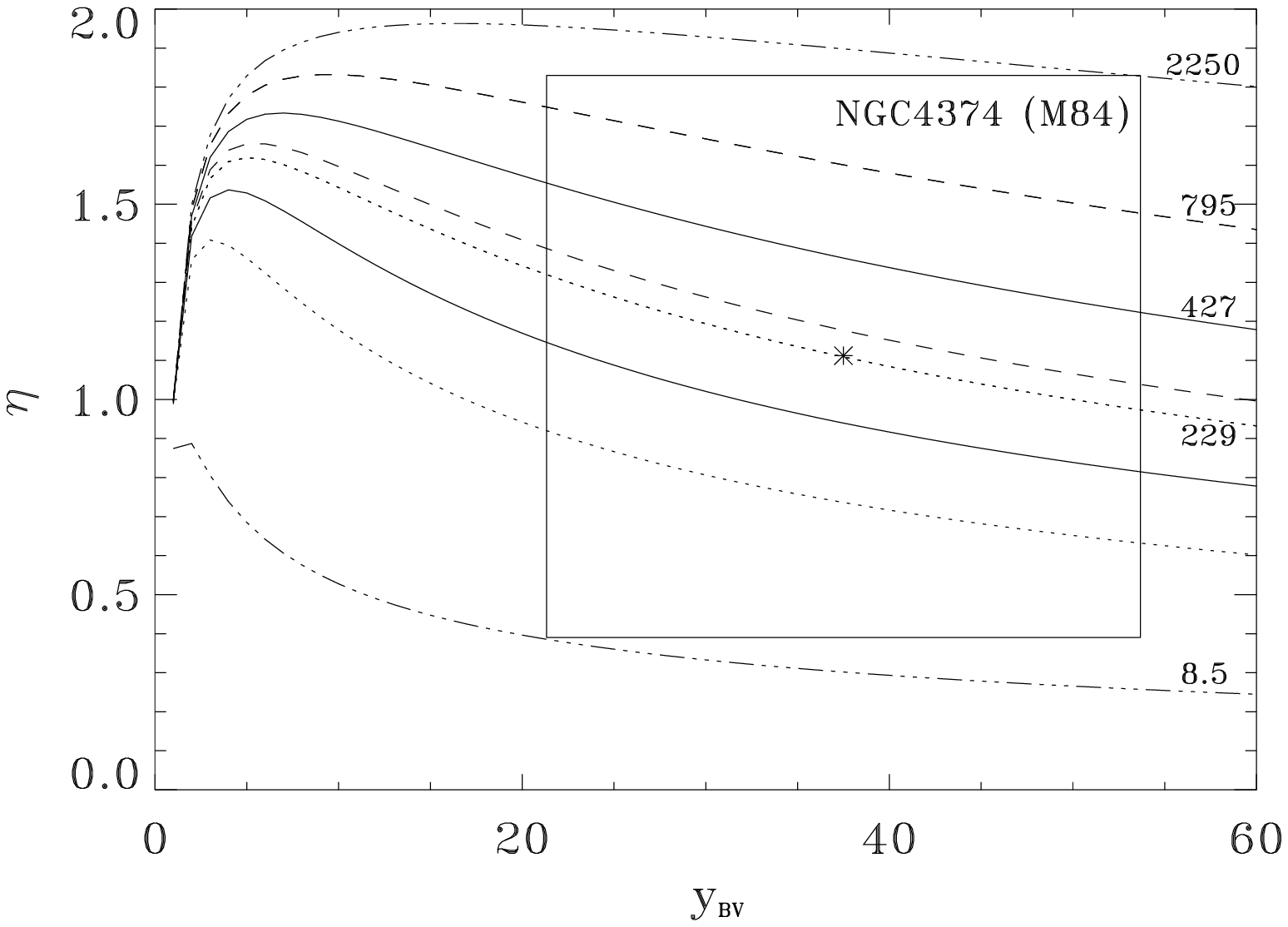}
\caption{Comparison between model predictions and data
from observations for NGC 4374 (M84).   The position
of the galaxy at the centre of the error box (deduced
from values listed in Tab.\,\ref{t:spec}) is marked by
an asterisk.   The value of the bulge to hole mass ratio,
$m_{\rm BH}$, can be read near the corresponding curve on
the right of the box error.   The value of the vortex to
hole mass ratio, $m_{\rm VH}$, is null in all cases except
lower full, dashed, and dotted curves, where $m_{\rm VH}=5$
and $m_{\rm BH}$ remains unchanged with respect to their
upper counterparts (value labelled therein).   The extreme
cases still compatible with the error box, are represented
by dot-dashed curves.   The centre of the error box is
fitted by $m_{\rm BH}=230$ (not represented).}
\label{f:N4374}    
\end{center}
\end{figure*}

\begin{figure*}
\begin{center}
\includegraphics[scale=0.8]{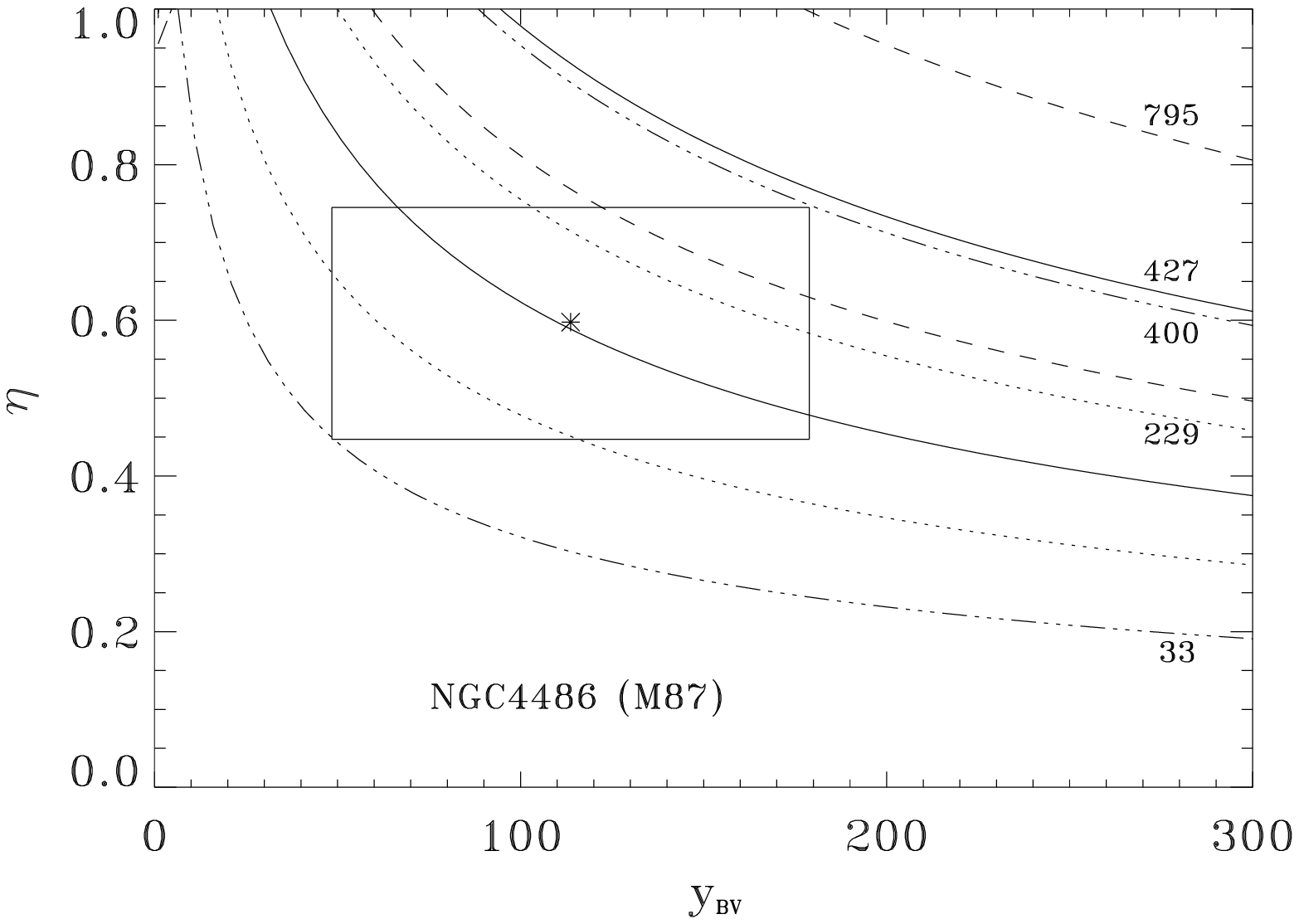}
\caption{Comparison between model predictions and data
from observations for NGC 4486 (M87).   The position
of the galaxy at the centre of the error box (deduced
from values listed in Tab.\,\ref{t:spec}) is marked by
an asterisk.   The value of the bulge to hole mass ratio,
$m_{\rm BH}$, can be read near the corresponding curve on
the right of the box error.   The value of the vortex to
hole mass ratio, $m_{\rm VH}$, is null in all cases except
lower full, dashed, and dotted curves, where $m_{\rm VH}=5$
and $m_{\rm BH}$ remains unchanged with respect to their
upper counterparts (value labelled therein).   The
extreme cases still compatible with the error box, are
represented by dot-dashed curves.   The centre of the error
box is fitted by $m_{\rm BH}=153$ (not represented).}
\label{f:N4486}    
\end{center}
\end{figure*}

The virial equations, Eqs.\,(\ref{eq:vspa})
and (\ref{eq:vspc}), provide the remaining
two unknowns: the fractional mass,
$m_{\rm BH}$,
and the fractional radius, $y_{\rm BV}$.
The resulting position of NGC 4374 and NGC 4486 on
the $({\sf O}y_{\rm BV}\eta)$ plane,
%
%
is $(y_{\rm BV},\eta)=(37.5,1.11)$ and
$(113.68,0.60)$, respectively.   The
dimensions of the error box can be
evaluated by use of standard quadratic
propagation formulae; the result is
$(2\sigma_{\rm y_{BN}},2\sigma_\eta)=
(32.34,1.44)$ and $(130,0.30)$,
respectively.   The comparison with
model predictions is shown in Figs.\,\ref
{f:N4374} and \ref{f:N4486} for NGC 4374
and NGC 4486, respectively.   The position
of each galaxy at the centre of the error
box is marked by an asterisk.   The value
of the fractional mass, $m_{\rm
BH}$, can be read near the corresponding
curve on the right of the error box.
The value of the fractional mass,
$m_{\rm VH}$, is null in all cases except
lower full, dashed, and dotted curves,
where $m_{\rm VH}=5$ and $m_{\rm BH}$
remains unchanged with respect to their
upper counterparts (value labelled therein).
The extreme cases still compatible
with the error box, are represented by
dot-dashed curves.    The centre of the
error box is fitted by $m_{\rm BH}=230$
(NGC 4371) and $m_{\rm BH}=153$ (NGC 4486);
the related curves are not plotted in
Figs.\,\ref{f:N4374} and \ref{f:N4486}, to
avoid confusion.

\subsection{Discussion}
\label{disc}

The application of the tensor virial theorem
for subsystems to hole + vortex + bulge
systems implies a number of restrictive
assumptions to gain intrinsic simplicity
related to analytical formulation.   This
is why the hole, the vortex, and the
inner bulge are modelled as homogeneous,
and the outer bulge as homeoidally striated.
Accordingly, the bulge mass, $M_{\rm B}$,
the inner bulge mass, $M_{\rm I}=M_{\rm B}
(R_{\rm V})$, and the fractional radius,
$y_{\rm BV}=R_{\rm B}/R_{\rm V}$,
are related by Eq.\,(\ref{eq:SSmas}).

For assigned fractional masses, $m_{\rm BH}$ and
$m_{\rm BV}$, the velocity ratio, $\eta=(\sigma_
{\rm B})_{33}/\{[(v_{\rm V})_{qq}]^2+[(\sigma_
{\rm V})_{qq}]^2\}^{1/2}$,
versus the fractional radius, $y_{\rm BV}=R_{\rm B}/
R_{\rm V}$, is represented by a selected curve on the
$({\sf O}y_{\rm BV}\eta)$ plane, as shown in Fig.\,\ref
{f:isfm}.   Interestingly, a fixed value of $\eta$
below the maximum corresponds to two different
configurations: a compact bulge on the left of
the maximum, and an extended bulge on the right.

On the other hand, fixed $m_{\rm BV}$
and $y_{\rm BV}$, imply larger $m_{\rm BH}$ passing
from bottom to top along a vertical line, $y_{\rm BV}
={\rm const}$,
as shown by the intersections of the above
mentioned vertical line with the corresponding curves,
see e.g., $y_{\rm BV}=50$ in Fig.\,\ref{f:isfm}.
Accordingly, larger $m_{\rm BH}$ imply larger values
of $\eta$ and vice versa.
Similar results hold for the fractional
mass, $m_{\rm BV}$, keeping $m_{\rm BH}$ fixed. 

Upper curves of each series are related to
massless vortexes, $m_{\rm VH}=0$.
Test particles moving on circular and
coplanar orbits, within a few hundred
parsecs from the central hole, may be
considered as massless vortexes.
Then the knowledge of orbital parameters
together with bulge effective radius and
velocity dispersion, defines the position
of the system on the $({\sf O}y_{\rm BV}
\eta)$ plane.    The isofractional mass
curve where the representative point lies,
in turn, defines the value of the fractional
mass, $m_{\rm BH}$.

The application to NGC 4374 and NGC 4486,
shown in Figs.\,\ref{f:N4374} and \ref
{f:N4486}, yields $8.5<m_{\rm BH}<2250$
and $33<m_{\rm BH}<400$, respectively.
The empirical $M_{\rm B}-M_{\rm H}$
correlation (Marconi and Hunt, 2003),
$229<m_{\rm BH}<795$, lies
within the uncertainty range for NGC 4374
but only partially for NGC 4486.   The
knowledge of the vortex density profile
would provide a better determination of
the mass-weighted rotation velocity,
$(V_{\rm V})_{qq}$, and then reduce the
error box.    In presence of a massive
vortex $(m_{\rm VH}=5)$, the hole mass
has to be reduced by a factor 2-3 to
get the fit.

The bulge mass can be estimated from
Figs.\,\ref{f:N4374} and \ref{f:N4486},
using hole mass values deduced from gas
dynamics listed in Tab.\,\ref{t:spec}.
The result is: $M_{\rm B}/{\rm M}_{10}=
34.35_{+550.65}^{-033.59}$ and $48.96_
{+115.04}^{-041.37}$ for NGC 4374 and
NGC 4486, respectively.

On the other
hand, the bulge mass may be inferred
from the knowledge of the mass to
luminosity ratio, the $I$-band
magnitude, and the distance modulus
(e.g., Caimmi and Valentinuzzi, 2008),
using recent data and related uncertainties
(Cappellari et al., 2006).   The result
is: $M_{\rm B}/{\rm M}_{10}=72.70\mp
67.50$ and $91.94\mp80.92$ for NGC 4374
and NGC 4486, respectively.

An alternative
approach consists in dealing with the structural
parameter estimate (Bender et al., 1992)
of $M_{\rm B}=(5/G)\sigma_o^2R_e$,
where $\sigma_o$ is the bulge
central velocity dispersion projected on
the line of sight, and $R_e$ is the bulge
effective radius.   An application to
NGC 4374 can be found in an earlier attempt
(Bower et al., 1998).    Using recent data
and related uncertainties (Cappellari et
al., 2006) yields $M_{\rm B}/{\rm M}_{10}=
72.70\mp22.62$ and $108.12\mp47.84$ for
NGC 4374 and NGC 4486, respectively.
The uncertainty should be
multiplied by $\zeta^\mp$, if $5\zeta^-<5<5
\zeta^+$ is the scatter of the above relation.

Finally, the substitution of hole mass
values listed in Tab.\,\ref{t:spec} into
the empirical $M_{\rm B}$-$M_{\rm H}$
correlation, $229<m_{\rm BH}<795$, with
a preferred value, $m_{\rm BH}=427$,
(Marconi and Hunt, 2003), yields
$M_{\rm B}/{\rm M}_{10}=64.05_{+142.65}^
{-043.44}$ and $136.64_{+189.31}^{-083.97}$
for NGC 4374 and NGC 4486, respectively.

\begin{figure*}
\begin{center}
\includegraphics[scale=0.8]{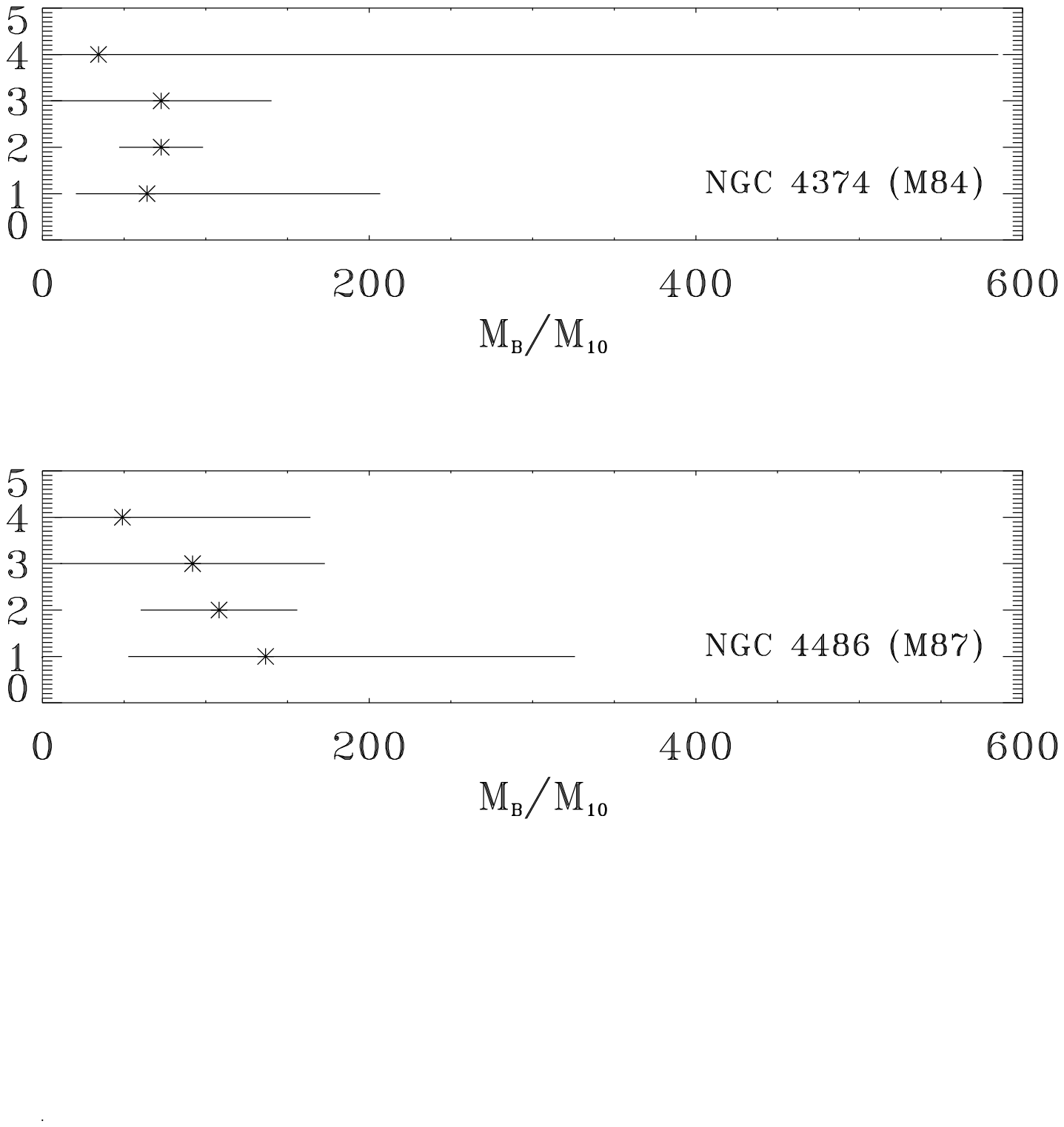}
\caption{The bulge mass deduced from different methods,
for NGC 4374 (M84, top) and NGC 4486 (M87, bottom).   
Different values in ordinate are deduced from: 1 - 
the empirical $M_{\rm B}$-$M_{\rm H}$
correlation, $229<m_{\rm BH}<795$, with
a preferred value, $m_{\rm BH}=427$,
(Marconi and Hunt, 2003), using hole mass
values listed in Tab.\,\ref{t:spec}; 2 - the structural
parameter estimate (Bender et al., 1992)
of $M_{\rm B}=(5/G)\sigma_o^2R_e$,
where $\sigma_o$ is the bulge
central velocity dispersion projected on
the line of sight, and $R_e$ is the bulge
effective radius (the upper and lower limit should be
multiplied by $\zeta^+$ and $\zeta^-$, respectively,
where $5\zeta^-<5<5
\zeta^+$ is the scatter of the above relation);
3 - the knowledge of the mass to
luminosity ratio, the $I$-band
magnitude, and the distance modulus
(e.g., Caimmi and Valentinuzzi, 2008),
using recent data and related uncertainties
(Cappellari et al., 2006); 4 - 
Figs.\,\ref{f:N4374} and \ref{f:N4486},
using hole mass values deduced from gas
dynamics listed in Tab.\,\ref{t:spec}.
Values of $M_{\rm B}/{\rm M}_{10}$ for
NGC 4374: 1 - $64.05_{+142.65}^
{-043.44}$; 2 - $72.70\mp22.62$; 3 - 
$72.70\mp67.50$; 4 - 
$34.35_{+550.65}^{-033.59}$.
Values of $M_{\rm B}/{\rm M}_{10}$ for
NGC 4486: 1 - $136.64_{+189.31}^{-083.97}$;
2 - $108.12\mp47.84$; 3 - $91.94\mp80.92$;
4 - $48.96_{+115.04}^{-041.37}$.}
\label{f:massa}    
\end{center}
\end{figure*}

A comparison between the above results is
shown in Fig.\,\ref{f:massa}.
Even if the current method is affected
by a large upper uncertainty for
NGC 4374, still it appears to be consistent
with the other ones.

In general, a selected point on the
$({\sf O}y_{\rm BV}\eta)$ plane is
crossed by infinite curves of the
kind plotted in Fig.\,\ref{f:isfm}, each
characterized
by a proper choice of fractional masses,
$(m_{\rm BV}, m_{\rm BH})$.   In the
limit of a massless vortex, $m_{\rm BV}
\to+\infty$, the model may be improved
by replacing the inner bulge radius,
$R_{\rm I}=R_{\rm V}$, with $R_{\rm I}=
R_{\rm H}$.   Accordingly, the bulge
resembles a truncated isothermal sphere,
where the central divergence of the
density profile is avoided by the presence
of a very small, homogeneous inner bulge.

\begin{figure*}
\begin{center}
\includegraphics[scale=0.8]{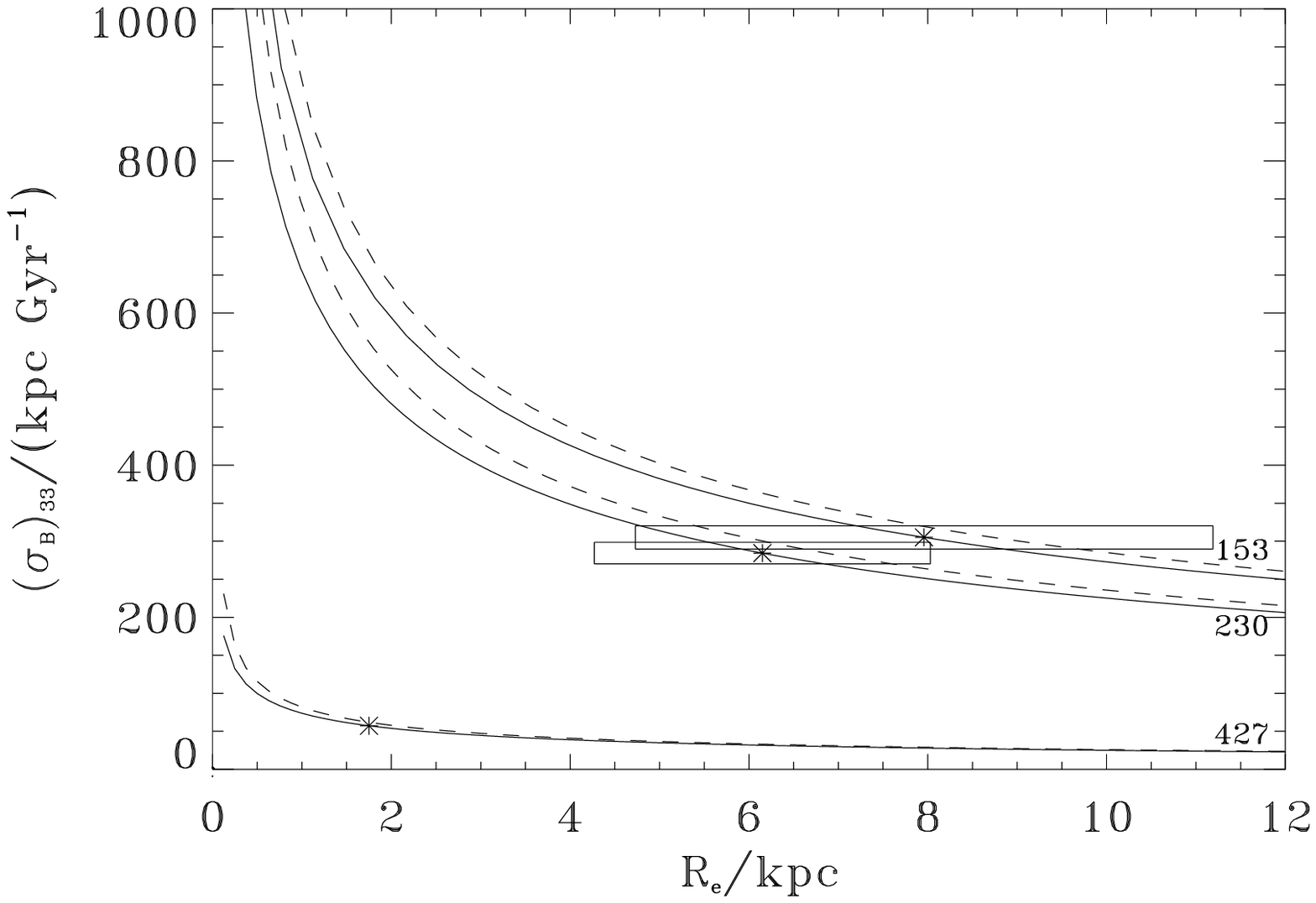}
\caption{The bulge velocity dispersion,
$(\sigma_{\rm B})_{33}$, as a function of the
bulge effective radius, $R_e$, in the
limit of a massless vortex, for $R_{\rm I}=
R_{\rm V}$ (full curves) and $R_{\rm I}
=R_{\rm H}$ (dashed curves), related to:
$(m_{\rm BH},R_{\rm V}/{\rm pc})=(427,
125),$ $(230, 164),$ $(153, 70)$, in
connection with Figs.\,\ref{f:isfm}, \ref
{f:N4374}, \ref{f:N4486}, respectively.
The position of the reference
configuration (Fig.\,\ref{f:isfm})
is marked by an asterisk.   The
position of NGC 4374 (Fig.\,\ref{f:N4374})
and NGC 4486 (Fig.\,\ref{f:N4486})
and the related uncertainties,
taken or deduced from recent
observations (Cappellari et al.,
2006), are marked by asterisks and
rectangles, respectively.}
\label{f:s33Re}    
\end{center}
\end{figure*}

The bulge velocity dispersion, $(\sigma_
{\rm B})_{33}$, as a function of the
bulge effective radius, $R_e$, in the
limit of a massless vortex, is plotted
in Fig.\,\ref{f:s33Re} for $R_{\rm I}=
R_{\rm V}$ (full curves) and $R_{\rm I}
=R_{\rm H}$ (dashed curves), related to:
$(m_{\rm BH},R_{\rm V}/{\rm pc})=(427,
125),$ $(230, 164),$ $(153, 70)$, in
connection with Figs.\,\ref{f:isfm}, \ref
{f:N4374}, \ref{f:N4486}, respectively.
The position of the reference
configuration (Fig.\,\ref{f:isfm})
is marked by an asterisk.   The
position of NGC 4374 and NGC 4486
and the related uncertainties,
taken or deduced from recent
observations (Cappellari et al.,
2006), are marked by asterisks and
rectangles, respectively.
It can be seen that the assumption
of a homogeneous inner bulge makes
bulge velocity dispersion systematically
decrease for fixed bulge effective radius,
but the error is comparable with
observational uncertainties.

The assumption of isotropic bulge
stress tensor is acceptable for
the special cases to which the
model has been applied.   More
specifically, both NGC 4374 and
NGC 4486 appear to be only slightly
flattened.   Even if the shape is
entirely due to velocity dispersion,
the stress tensor is expected to be
moderately anisotropic provided the
system is seen edge-on, and can be
considered isotropic to a first
extent.   It is the case for
NGC 4486, but not for NGC 4374
(Bender et al., 1992).    If,
on the other hand, the system is seen
head-on, the line of sight coincides
with the polar axis and what is
deduced from observations is related to
$(\sigma_{\rm B})_{33}$, regardless
of the properties of the stress tensor.
Strictly speaking, the degree
of anisotropy and the inclination
angle should be specified, in order
to express the bulge velocity
dispersion component along the
polar axis, $(\sigma_{\rm B})_{33}$,
in terms of the luminosity-weighted
second moment of the line-of-sight
velocity distribution within the
bulge effective radius, $R_e$.

\section{Conclusion}\label{conc}

The tensor virial theorem for subsystems
has been formulated for three-component
systems and further effort has been devoted
to a special case where the inner subsystems
and the central region of the outer one are
homogeneous, the last surrounded by an
isothermal homeoid.   The virial equations
have explicitly been written under the
additional restrictions: (i) similar and
similarly placed inner subsystems, and
(ii) spherical outer subsystem.   An
application has been made to hole +
vortex + bulge systems, in the limit
of flattened inner subsystems, which
implies three virial equations in three
unknowns.

Using the Faber-Jackson relation,
$(R_{\rm B})_e\propto\sigma_0^2$,
the standard $M_{\rm H}$-$\sigma_0$
form $(M_{\rm H}\propto\sigma_0^4)$
has been deduced from qualitative
considerations.   The projected
bulge velocity dispersion to
projected vortex velocity
ratio, $\eta=(\sigma_{\rm B})_{33}/\{
[(v_{\rm V})_{qq}]^2+[(\sigma_{\rm V})_
{qq}]^2\}^{1/2}$, as a
function of the fractional radius,
$y_{\rm BV}=R_{\rm B}/R_{\rm V}$,
and the fractional masses,
$m_{\rm BH}=M_{\rm B}/M_{\rm H}$ and
$m_{\rm BV}=M_{\rm B}/M_{\rm V}$,
has been studied in the range of
interest, $0\le m_{\rm VH}=
M_{\rm V}/M_{\rm H}\le5$ (Escala,
2006) and $229\le m_{\rm BH}\le795$
(Marconi and Hunt, 2003), consistent
with observations.

The related curves have been shown to be
similar to Maxwell velocity distributions,
which implies a fixed value of $\eta$
below the maximum corresponds to two
different configurations: a compact bulge
on the left of the maximum, and an
extended bulge on the right.    All curves
have been seen to lie very close one to
the other on the left of the maximum, and
parallel one to the other on the right.

On the other hand, fixed $m_{\rm BH}$ or
$m_{\rm BV}$, and $y_{\rm BV}$, have
been found to imply more massive bulges
passing from bottom to top along a vertical
line on the $({\sf O}y_{\rm BV}\eta)$ plane,
and vice versa.    The model has been applied
to NGC 4374 and NGC 4486, taking the fractional
mass, $m_{\rm BH}$, and the fractional
radius, $y_{\rm BV}$, as
unknowns, and the bulge mass has been inferred
and compared with results from different
methods.    In presence of a massive
vortex $(m_{\rm VH}=5)$, the fit has been
provided by hole mass 2-3 times lower with
respect to the case of a massless vortex.
Finally, it has been shown that
the assumptions of homogeneous inner bulge
and isotropic stress tensor
hold to an acceptable extent, at least in
the special cases taken into consideration.

\section*{Acknowledgements}
Thanks are due to S. Masiero for fruitful
discussions.

\appendix
\section*{Appendix}

\section{Ellipsoid shape factors: special cases}
\label{a:sfa}

Let $a_1$, $a_2$, $a_3$, $a_1\ge a_2\ge a_3$, be
semiaxes of a generic ellipsoid, and $\epsilon_
{pq}=a_p/a_q$ related axis ratios.   The shape
factors, $B_p$ and $B_{pr}$, defined by Eqs.\,(\ref
{eq:EVSsb}) and (\ref{eq:EXYth}), respectively,
in terms of their usual counterparts (e.g.,
Chandrasekhar, 1969, Chap.\,3, \S21; Caimmi,
1991, 1995), read:
\begin{lefteqnarray}
\label{eq:ABp}
&& B_p=\epsilon_{p2}\epsilon_{p3}A_p~~; \\
\label{eq:ABpr}
&& B_{pr}=\epsilon_{p2}\epsilon_{p3}A_{pr}a_p^2~~; \\
\label{eq:ABr}
&& \frac{B_{pr}}{B_p}=\frac{A_{pr}a_p^2}{A_p}~~;
\end{lefteqnarray}
and a number of special cases are listed
in Tab.\,\ref{t:Bsha} for limiting
configurations.

\begin{table}
\begin{tabular}{cccccc}
\hline
\hline
\multicolumn{1}{c}{} &
\multicolumn{1}{c}{oblong} &
\multicolumn{1}{c}{flat}
&
\multicolumn{1}{c}{prolate} &
\multicolumn{1}{c}{oblate} &
\multicolumn{1}{c}{spherical} \\
\hline
$\epsilon_{21}$ & 0         & $\epsilon_f$ & $\epsilon_p$          & 1                   & 1   \\
$\epsilon_{31}$ & 0         & 0            & $\epsilon_p$          & $\epsilon_o$        & 1   \\
$B_1$           & $+\infty$ & $\pi/2$      & $\gamma/\epsilon_p^2$ & $\alpha/\epsilon_o$ & 2/3 \\
$B_2$           & 1         & $\pi/2$      & $\alpha$              & $\alpha/\epsilon_o$ & 2/3 \\
$B_3$           & 1         & 0            & $\alpha$              & $\epsilon_o\gamma$  & 2/3 \\
$B_{11}$        & $+\infty$ & $3\pi/8$     & $w_3/\epsilon_p^2$    & $w_1/\epsilon_o$    & 2/5 \\
$B_{12}$        & $+\infty$ & $3\pi/8$     & $w_2/\epsilon_p^4$    & $w_1/\epsilon_o$    & 2/5 \\
$B_{13}$        & $+\infty$ & $+\infty$    & $w_2/\epsilon_p^4$    & $w_2/\epsilon_o$    & 2/5 \\
$B_{21}$        & 0         & $3\pi/8$     & $w_2$                 & $w_1/\epsilon_o$    & 2/5 \\
$B_{22}$        & 1/2       & $3\pi/8$     & $w_1$                 & $w_1/\epsilon_o$    & 2/5 \\
$B_{23}$        & 1/2       & $+\infty$    & $w_1$                 & $w_2/\epsilon_o$    & 2/5 \\
$B_{31}$        & 0         & 0            & $w_2$                 & $\epsilon_o^3w_2$   & 2/5 \\
$B_{32}$        & 1/2       & 0            & $w_1$                 & $\epsilon_o^3w_2$   & 2/5 \\
$B_{33}$        & 1/2       & 0            & $w_1$                 & $\epsilon_ow_3$     & 2/5 \\
\hline\hline
\end{tabular}
\caption{Values of ellipsoid shape factors,
$B_p$ and $B_{pr}$, related to limiting
configurations defined by the values of
the axis ratios, $\epsilon_{21}$ and
$\epsilon_{31}$.   For oblong configurations,
$B_{1}$ and $B_{11}$ diverge as $\epsilon^2\gamma$,
$B_{12}$ and $B_{13}$ as $\epsilon^2$.
For flat configurations, $B_{13}$ and $B_
{23}$ diverge as $\epsilon^{-1}$.}
\label{t:Bsha}
\end{table}
The range of validity of the related independent
variables, $\epsilon_f$, $\epsilon_p$,
$\epsilon_o$, and $\epsilon$,
is:
\begin{leftsubeqnarray}
\slabel{eq:epa}
&& 0<\epsilon_f\le1~~;\qquad0<\epsilon_p<1~~;\qquad0<\epsilon_o<1~~; \\
\slabel{eq:epb}
&& \epsilon=\epsilon_{31},\qquad0<\epsilon<1,\qquad{\rm oblate}~~;
\\
\slabel{eq:epc}
&& \epsilon=\epsilon_{21}^{-1}=\epsilon_{31}^{-1},\qquad1<\epsilon<
+\infty,\qquad{\rm prolate}~~;
\label{seq:ep}
\end{leftsubeqnarray}
and the explicit expression of the functions,
$\alpha$, $\gamma$, $w_1$, $w_2$, $w_3$, reads:
\begin{lefteqnarray}
\label{eq:alpha}
&& \alpha=\cases{
\displayfrac{\epsilon}{1-\epsilon^2}\left[\displayfrac{\arcsin\sqrt
{1-\epsilon^2}}{\sqrt{1-\epsilon^2}}-\epsilon\right]~~;  & oblate; \cr
\displayfrac{\epsilon}{\epsilon^2-1}\left[\epsilon-\displayfrac{\arcsinh\sqrt
{\epsilon^2-1}}{\sqrt{\epsilon^2-1}}\right]~~;  & prolate; \cr} \\
\label{eq:gamma}
&& \gamma=\cases{
\displayfrac2{1-\epsilon^2}\left[1-\epsilon\displayfrac{\arcsin\sqrt
{1-\epsilon^2}}{\sqrt{1-\epsilon^2}}\right]~~;  & oblate; \cr
\displayfrac2{\epsilon^2-1}\left[\epsilon\displayfrac{\arcsinh\sqrt
{\epsilon^2-1}}{\sqrt{\epsilon^2-1}}-1\right]~~;  & prolate; \cr} \\
\label{eq:w} 
&& w_1=\frac14\left(3\alpha-\epsilon^2w_2\right)~~;\qquad w_2=\frac
{\gamma-\alpha}{1-\epsilon^2}~~;\qquad w_3=\frac23\left(1-\epsilon^2w_2
\right)~~;\qquad
\end{lefteqnarray}
where, in addition (e.g., Caimmi, 1991):
\begin{lefteqnarray}
\label{eq:m1}
&& 2\alpha+\gamma=2~~;\quad\lim_{\epsilon\to1}\alpha=\lim_{\epsilon\to1}
\gamma=\frac23~~;\quad
\lim_{\epsilon\to0}\alpha=0~~;\quad
\lim_{\epsilon\to0}\gamma=2~~; \\
\label{eq:m2}
&& \lim_{\epsilon\to0}\frac\alpha\epsilon=\frac\pi2~~;\qquad
\lim_{\epsilon\to+\infty}\alpha=1~~;\qquad\lim_{\epsilon\to+\infty}\gamma=0~~;
\qquad \lim_{\epsilon\to+\infty}\epsilon\gamma=0~~;\qquad \\
\label{eq:m3}
&& \lim_{\epsilon\to+\infty}\epsilon^2\gamma=+\infty~~;\qquad
\lim_{\epsilon\to1^-}w_2=\lim_{\epsilon\to1^+}w_2=\frac25~~.
\end{lefteqnarray}

\end{document}